\documentclass[12pt]{article}

\usepackage{latexsym,amsmath}

\DeclareFontFamily{OT1}{rsfs10}{}
\DeclareFontShape{OT1}{rsfs10}{m}{n}{ <-> rsfs10 }{}
\DeclareMathAlphabet{\mathscript}{OT1}{rsfs10}{m}{n}

\def\bcite{\@ifnextchar [{\@tempswatrue\@bcitex}{\@tempswafalse\@bcitex[]}}
\def\@bcitex[#1]#2{\if@filesw\immediate\write\@auxout{\string\citation{#2}}\fi
  \let\@bcitea\@empty
  \@bcite{\@for\@bciteb:=#2\do
    {\@bcitea\def\@bcitea{,\penalty\@m\ }%
     \def\@tempa##1##2\@nil{\edef\@bciteb{\if##1\space##2\else##1##2\fi}}%
     \expandafter\@tempa\@bciteb\@nil
     \@ifundefined{b@\@bciteb}{{\reset@font\bf ?}\@warning
       {Citation `\@bciteb' on page \thepage \space undefined}}%
     \hbox{\csname b@\@bciteb\endcsname}}}{#1}}
\def\@bcite#1#2{{#1\if@tempswa , #2\fi}}

\newcommand{\ns}{\normalsize}

\newcommand{\be}{\begin{equation}}
\newcommand{\ee}{\end{equation}}
\newcommand{\nn}{\nonumber}
\newcommand{\bea}{\begin{eqnarray}}
\newcommand{\eea}{\end{eqnarray}}

\newcommand{\tr}{\textrm{tr}}

\def\a{\alpha}
\def\b{\beta}
\def\g{\gamma}
\def\c{\chi}
\def\d{\delta}
\def\e{\epsilon}

\def\z{\psi}

\def\k{\kappa}
\def\l{\lambda}
\def\m{\mu}
\def\n{\nu}
\def\o{\omega}
\def\p{\pi}

\def\r{\rho}
\def\s{\sigma}

\def\t{\tau}

\def\x{\xi}
\def\z{\zeta}
\def\w{\wedge}
\def\D{\Delta}

\def\G{\Gamma}
\def\J{\Psi}

\def\O{\Omega}

\def\cA{{\cal A}}
\def\cF{{\cal F}}
\def\cM{{\cal M}}

\begin{document}


\begin{titlepage}

\title{
   \vspace{-4em}             
   \hfill{\ns UPR-825T, OUTP-98-85P} 
   \vfill
   {\LARGE Cosmology and Heterotic M--Theory in Five--Dimensions}}
\author{
   {\ns Andr\'e Lukas$^1$, Burt A.~Ovrut$^2$
      \setcounter{footnote}{1}\thanks{Lectures presented at the
      Advanced School on Cosmology and Particle Physics, June 1998,
      Peniscola, Spain.}~~
      \setcounter{footnote}{2}\thanks{Supported in part by a Senior 
          Alexander von Humboldt Award.}~~
      and Daniel Waldram$^3$} \\
   {\it\ns $^1$Department of Physics, Oxford University} \\[-0.5em]
      {\it\ns 1 Keble Road, Oxford OX1 3NP, United Kingdom} \\   
   {\it\ns $^2$Department of Physics, University of Pennsylvania} \\[-0.5em]
      {\it\ns Philadelphia, PA 19104--6396, USA}\\
   {\it\ns $^3$Department of Physics, Joseph Henry Laboratories,} \\[-0.5em]
      {\it\ns Princeton University, Princeton, NJ 08544, USA}}
\date{}

\maketitle
\vspace{-1em}

\begin{abstract}
In these lectures, we present cosmological vacuum solutions of 
Ho\v rava-Witten theory and discuss their physical properties. We begin by 
deriving the five--dimensional effective action of strongly coupled
heterotic string theory
by performing a reduction, on a Calabi--Yau three--fold, of M--theory
on $S^1/Z_2$. The effective theory is shown to be a gauged version of
five--dimensional $N=1$ supergravity coupled, for simplicity, to the
universal hypermultiplet and four--dimensional boundary theories with
gauge and universal gauge matter fields. The static vacuum of the
theory is a pair of BPS three--brane domain walls. We show that this
five--dimensional theory, together with the domain wall vacuum
solution, provides the correct starting point for early universe
cosmology in Ho\v rava-Witten theory. Relevant cosmological solutions
are those associated with the BPS domain wall vacuum. Such solutions
must be inhomogeneous, depending on the orbifold coordinate as well as
on time. We present two examples of this new type of cosmological
solution, obtained by separation of variables. The first example
represents the analog of a rolling radii solution with the radii
specifying the geometry of the domain wall pair. This is generalized
in the second example to include a nontrivial Ramond-Ramond scalar. 

\end{abstract}

\thispagestyle{empty} 

\end{titlepage}


\section*{Introduction:}


It has been known for a long time that there are five consistent superstring
theories, all defined in ten--dimensional spacetime. These are the two $N=2$
supersymmetric string theories called Type IIA and IIB and three $N=1$
superstrings called Type I and heterotic $SO(32)$ and heterotic
$E_{8} \times E_{8}$. A few years ago, it was demonstrated that these apparently
different theories are, in fact, all related to each other through an
intricate web of so--called duality transformations. Furthermore, it became
clear that these five superstring theories are simply part of a much larger
moduli space associated with a more fundamental and unique theory, termed
$M$-theory. Interestingly, it was shown that there is yet another region of
the moduli space of $M$-theory that corresponds to the theory of
eleven--dimensional
supergravitation. It follows that $M$-theory cannot be a superstring theory,
but, to date, the exact nature of $M$-theory remains unknown. Be that as it
may, $M$-theory does emphasize the fundamental importance of
eleven--dimensional supergravity as a starting point for exploring aspects of
cosmology and particle physics at energies near or below the
eleven-dimensional Planck scale. In these lectures, we will discuss what is
presently known about the exact structure of $M$-theory at low energy and
present the first results on the cosmological solutions and behaviour of this
theory. We refer the reader to reference~\cite{sen} for a detailed discussion of
duality in superstring theory and to references therein.

The strongly coupled $E_8\times E_8$ heterotic string has been
identified as the eleven-dimensional limit of M-theory
compactified on an $S^1/Z_2$ orbifold with a set of $E_8$ gauge supermultiplets
on each ten-dimensional orbifold fixed plane~\cite{hw1,hw2}. Witten
has shown that there exists a consistent compactification of this M-theory
limit on a deformed Calabi-Yau three--fold, leading to a supersymmetric $N=1$
theory in four dimensions~\cite{w}. Matching at tree level to the
phenomenological gravitational and grand-unified gauge couplings~\cite{w,bd}, one
finds that the orbifold must be larger than the Calabi-Yau radius, which is of
the order of the inverse gauge coupling unification scale, about $10^{16}$ GeV. 
Typically, one finds that the orbifold radius can be chosen to be from a
factor of four or five times the inverse unification scale to as large as 
inverse $10^{14}$
GeV, depending on circumstances. This suggests that there
is a substantial regime where the universe appears five-dimensional. It is
important, therefore,  to determine the five-dimensional effective
action describing heterotic M--theory in this regime. This theory
constitutes a new setting for early universe 
cosmology, which has traditionally been studied in the framework of
the four-dimensional effective action.

In previous papers~\cite{losw,losw1}, we derived this five--dimensional
effective theory by
directly reducing Ho\v rava--Witten theory on a Calabi--Yau three--fold.
We showed that a non--zero mode of the antisymmetric tensor field
strength has to be included for a consistent reduction from eleven to
five dimensions and that the correct five--dimensional effective
theory of the strongly coupled heterotic string is given by a gauged
version of five--dimensional supergravity. We explicitly
included all $(1,1)$ moduli vector superfields and the universal
hypermultiplet, as well as indicating how the remaining $(2,1)$ moduli could
be incorporated. We also discussed the four--dimensional boundary
gauge and matter supermultiplets and derived their effective action. 
In these lectures, we present the effective five-dimensional theory for
the universal bulk fields only; that is, the gravity supermultiplet and the
universal hypermultiplet. We show that the consistent reduction
from eleven to five dimensions on a Calabi-Yau manifold requires the inclusion
of non-zero values of the four-form field strength in the internal Calabi-Yau
three--fold directions. This leads to a gauged five-dimensional supergravity action
with a potential term. More precisely,
given the universal hypermultiplet coset manifold~\cite{quat}
$\cM_Q=SU(2,1)/SU(2)\times U(1)$, we find that a subgroup
$U(1)\subset SU(2)\times U(1)$ is gauged, with the vector field in
the gravity supermultiplet as the corresponding gauge boson. Due to the
potential, flat space is not a solution of this five-dimensional
theory. However, the
equations of motion do admit as a solution a pair of 
three-branes that preserves
half of the remaining $d=5$ supersymmetries.
This solution is supported by source terms on the fixed orbifold planes
of the five-dimensional space. This pair of BPS three-branes constitutes the
static ``vacuum'' of the five-dimensional theory and it is the appropriate
background for a further reduction to four-dimensional $N=1$ supergravity
theories. In such a reduction, four-dimensional space-time becomes
identified with the three-brane worldvolume. We will show that
the linearized version of this three-brane corresponds to Witten's
``deformed'' Calabi-Yau solution, which was constructed only to first
non-trivial order in powers of the eleven-dimensional Newton constant.
Subsequent work discussing Ho\v rava-Witten and related theories from the
point of view of five--dimensional supergravity with boundaries can be found
in ~\cite{lalak,lisa}.

It is clear from the above discussion that
cosmology in effective Ho\v rava-Witten theory should be studied in the 
five-dimensional context and will, generically, be 
inhomogeneous in the extra spatial
dimension. What should realistic theories look like? In the ideal case,
one would have a situation in which the internal six-dimensional
Calabi-Yau space and the orbifold evolve for a period of time
and then settle down to their ``phenomenological'' values while the
three non--compact dimensions continue to expand. Then, for late time,
when all physical scales are much larger than the orbifold size, the
theory is effectively four-dimensional and should, in the ``static''
limit, provide a realistic supergravity model of particle physics. As
we have argued above, such realistic supergravity models originate
from a reduction of the five-dimensional theory on its BPS domain wall
vacuum state. Hence, in the ``static'' limit at late time, realistic
cosmological solutions should reduce to the BPS domain wall vacuum or, perhaps, a
modification thereof that incorporates spontaneous breaking of the remaining
four-dimensional $N=1$ supersymmetry. Consequently, one is forced to
look for solutions which depend on the orbifold coordinate as well as
on time. In a previous paper~\cite{us1}, we presented examples of
such cosmological solutions in five-dimensional heterotic M-theory. In these
lectures, we review these solutions and
illustrate some of the characteristic cosmological features of the
theory.

In earlier work~\cite{lowc1,lowc2}, we showed how a general class of
cosmological solutions, that is, time-dependent solutions of the
equations of motion that are homogeneous and isotropic in our physical
$d=3$ subspace, can be obtained in both superstring theories and
M-theory defined in spacetimes without boundary. Loosely
speaking, we showed that a cosmological solution could be obtained
from any p-brane or D-brane by inverting the roles of the time and
``radial'' spatial coordinate. This method will clearly continue to
work in Ho\v rava-Witten theory as long as one exchanges time with a
radial coordinate not aligned in the orbifold direction. An example of
this in eleven-dimensions, based on the solution of~\cite{llo}, has
been given in~\cite{benakli}. It can not, however, be applied to the
fundamental domain wall since its radial direction coincides with the
orbifold coordinate. This coordinate is bounded and cannot be turned
into time. Instead, the domain wall itself should be made time dependent
thereby leading to solutions that depend on both time and the orbifold
coordinate. As a result, we have to deal with coupled partial differential
equations, but, under certain constraints, these can by solved by separation
of variables, though the equations remain non-linear. Essentially, we are
allowing the moduli describing the geometry of the domain wall and the
excitations of other five-dimensional fields, to become
time-dependent. Technically, we will simply take the usual Ans\"atze
for the five-dimensional fields, but now allow the functions to depend
on both the time and radial coordinates. We will further demand
that these functions each factor into a purely time dependent piece
times a purely radial dependent piece. This is not, in general,
sufficient to separate the equations of motion. However, we will show
that, subject to certain constraints, separation of variables is
achieved. We can solve these separated equations and find new,
cosmologically relevant solutions. In these lectures, we will restrict our
attention to two examples representing cosmological extensions of the
pure BPS pair of three-branes. 

The first example is simply the domain wall itself with two of its three
moduli made time-dependent. We show that separation of variables occurs
in this case. It turns out that these moduli behave like
``rolling radii''~\cite{mueller}, which constitute fundamental cosmological
solutions in weakly coupled string theory. Unlike those rolling radii
which represent scale factors of homogeneous, isotropic spaces, here they
measure the separation of the two walls of the three-brane and its
worldvolume size (which, at the same time, is the size of ``our''
three-dimensional universe). We have, therefore,  a time-dependent
domain wall pair with its shape staying rigid but its size and separation
evolving like rolling radii. 

For the second example, we consider a similar setting as
for the first but, in addition, we allow a
nonvanishing Ramond-Ramond scalar. This terminology
is perhaps a little misleading, but relates to the fact that the
scalar would be a Type II Ramond-Ramond field in the case where the
orbifold was replaced by a circle. This makes connection with
Type II cosmologies with non-trivial Ramond-Ramond fields discussed
in~\cite{lowc1,lowc2}. Separation of variables occurs for a
specific time-independent form of this scalar. The
orbifold-dependent part then coincides with the domain wall with, however, the
addition of the Ramond-Ramond scalar. This non--vanishing value of the
scalar breaks supersymmetry even in the static limit.
We find that the time-dependent part of the equations fits into the
general scheme of M-theory cosmological solutions with form fields as
presented in ref.~\cite{lowc1,lowc2}. Applying the results of these
papers, the domain wall moduli are found to behave like rolling radii
asymptotically for early and late times. The evolution rates in these
asymptotic regions are, however, different and the transitions between them can
be attributed to the nontrivial Ramond-Ramond scalar. 

Let us summarize our conventions. We will
consider eleven-dimensional spacetime compactified on a Calabi-Yau space $X$,
with the subsequent reduction down to four dimensions effectively provided by
a double-domain-wall background, corresponding to an $S^1/Z_2$ orbifold. We
use coordinates $x^{I}$ with indices $I,J,K,\ldots = 0,\ldots ,9,11$ to
parameterize the full eleven-dimensional space $M_{11}$. Throughout these
lectures, when we refer to orbifolds, we will work in the ``upstairs'' picture
with the orbifold $S^1/Z_2$ in the $x^{11}$-direction. We choose the range
$x^{11}\in [-\pi\rho ,\pi\rho ]$ with the endpoints being identified. The
$Z_2$ orbifold symmetry acts as $x^{11}\rightarrow -x^{11}$. Then there exist
two ten-dimensional hyperplanes fixed under the $Z_2$ symmetry which we
denote by $M_{10}^{(i)}$, $i=1,2$. Locally, they are specified by the
conditions $x^{11}=0,\pi\rho$. Barred indices
$\bar{I},\bar{J},\bar{K},\ldots = 0,\ldots ,9$ are used for the
ten-dimensional space orthogonal to the orbifold. 
Upon reduction on the Calabi-Yau space we have a five-dimensional spacetime
$M_5$ labeled by indices $\a ,\b ,\g ,\ldots  = 0,\ldots ,3,11$. The
orbifold fixed planes become four-dimensional with indices
$\m,\n,\rho,\ldots = 0,\ldots ,3$. We use indices $A,B,C,\ldots =
4,\ldots 9$ for the Calabi-Yau space. 
The eleven-dimensional Dirac-matrices $\G^I$ with $\{\G^I,\G^J\}=2g^{IJ}$
are decomposed as $\G^I = \{\g^\a\otimes\l ,{\bf 1}\otimes\l^A\}$
where $\g^\a$ and $\l^A$ are the five- and six-dimensional Dirac
matrices, respectively. Here, $\l$ is the chiral projection matrix in
six dimensions with $\l^2=1$. Spinors in eleven dimensions will be
Majorana spinors with 32 real components throughout the paper. In five
dimensions we use symplectic-real spinors~\cite{c0} $\psi^i$ where
$i=1,2$ is an $SU(2)$ index, corresponding to the automorphism
group of the $N=1$ supersymmetry algebra in five dimensions. We will
follow the conventions given in~\cite{GST1}.
Fields will be required to have a definite behaviour under the $Z_2$
orbifold symmetry in $D=11$. We demand a bosonic field $\Phi$ to be
even or odd; that is, $\Phi (x^{11})=\pm\Phi (-x^{11})$. For a spinor
$\Psi$ the condition is $\G_{11}\Psi (-x^{11})=\pm \Psi (x^{11})$ so that
the projection to one of the orbifold planes leads to a
ten-dimensional Majorana-Weyl spinor with definite
chirality. Similarly, in five dimensions, bosonic fields will be
either even or odd. We can choose a basis for the $SU(2)$ automorphism
group such that symplectic-real spinors $\psi^i$ 
satisfy the constraint
$\g_{11}\psi^i(-x^{11})=(\t_3)^i_j\psi^j(x^{11})$ where $\t_a$ are the
Pauli spin matrices, so $\t_3=\mbox{diag}(1,-1)$.


\section*{Lecture 1:}


In this first lecture, we will discuss the Ho\v{r}ava--Witten theory of
eleven-dimensional supergravity compactified on an $S^{1}/Z_{2}$ orbifold. We
then, using the standard embedding of the spin connection into the gauge
connection, discuss the compactification of this theory to four--dimensions so
that a single $N=1$ supersymmetry is preserved. The constraint that $N=1$
supersymmetry be preserved causes a ``deformation'' of the background
spacetime. We show, in detail, that mathematical simplification and physical
clarity can be achieved by first compactifying Ho\v{r}ava--Witten on a
Calabi--Yau three-fold to five--dimensions. We construct the five--dimensional
effective theory in detail and show that it is a new, gauged form of
supergravity that does not admit flat spacetime as its static vacuum. We end
the lecture by discussing the supersymmetry variations of the gravitino and
hypermultiplet fermions.

\section*{The strongly coupled heterotic string and Calabi-Yau solutions}

To set the scene for our later discussion, we will now briefly review the
effective description of strongly coupled heterotic string theory as
eleven-dimensional supergravity with boundaries given by Ho\v{r}ava and
Witten~\cite{hw1,hw2}. In addition, we present, in a simple form, the
solutions of this theory~\cite{w} appropriate for a reduction to $N=1$
theories in four dimensions using the explicit form of these solutions given
in ref.~\cite{low1}.

The bosonic part of the action is of the form
\begin{equation}
\label{action}
   S = S_{\rm SG}+S_{\rm YM}
\end{equation}
where $S_{\rm SG}$ is the familiar eleven-dimensional supergravity
\begin{equation}
 S_{\rm SG} = -\frac{1}{2\k^2}\int_{M^{11}}\sqrt{-g}\left[ 
                    R+\frac{1}{24}G_{IJKL}G^{IJKL} 
           +\frac{\sqrt{2}}{1728}\e^{I_1...I_{11}}
               C_{I_1I_2I_3}G_{I_4...I_7}G_{I_8...I_{11}} \right]
\label{SSG}
\end{equation}
and $S_{\rm YM}$ are the two $E_8$ Yang-Mills theories on the orbifold planes
explicitly given by~\footnote{We note that there is a debate in the
literature about the precise value of the Yang-Mills coupling constant
in terms of $\k$. While we quote the original value~\cite{hw2,deA}, the
value found in ref.~\cite{conrad} is smaller.
In the second case, the coefficients in the Yang-Mills action~\eqref{SYM} 
and the Bianchi identity~\eqref{Bianchi} should both be multiplied by 
$2^{-1/3}$. This potential factor will not be essential in the following 
discussion as it will simply lead to a redefinition of the five-dimensional 
coupling constants. We will comment on this point later on.}
\begin{multline}
   S_{\rm YM} = - \frac{1}{8\pi\k^2}\left(\frac{\k}{4\pi}\right)^{2/3}
           \int_{M_{10}^{(1)}}\sqrt{-g}\;\left\{
           \tr(F^{(1)})^2 - \frac{1}{2}\tr R^2\right\} \\ {}
        - \frac{1}{8\pi\k^2}\left(\frac{\k}{4\pi}\right)^{2/3}
           \int_{M_{10}^{(2)}}\sqrt{-g}\;\left\{
           \tr(F^{(2)})^2 - \frac{1}{2}\tr R^2\right\}\; .
\label{SYM}
\end{multline}
Here $F_{\bar{I}\bar{J}}^{(i)}$ are the two $E_8$ gauge field strengths and
$C_{IJK}$ is the 3-form with field strength
$G_{IJKL}=24\,\partial_{[I}C_{JKL]}$. In order for the above 
theory to be supersymmetric as well as anomaly free, the Bianchi
identity for $G$ should receive a correction such that
\begin{equation}
 (dG)_{11\bar{I}\bar{J}\bar{K}\bar{L}} = 
    -\frac{1}{2\sqrt{2}\pi}
    \left(\frac{\k}{4\pi}\right)^{2/3} \left\{ 
       J^{(1)}\d (x^{11}) + J^{(2)}\d (x^{11}-\pi\r )
       \right\}_{\bar{I}\bar{J}\bar{K}\bar{L}} \label{Bianchi}
\end{equation}
where the sources are given by 
\begin{equation}
 J^{(i)}
    = {\rm tr}F^{(i)}\wedge F^{(i)} 
      - \frac{1}{2}{\rm tr}R\wedge R \; .\label{J}
\end{equation}
Under the $Z_2$ orbifold symmetry, the field components $g_{\bar{I}\bar{J}}$,
$g_{11,11}$, $C_{\bar{I}\bar{J}11}$ are even, while $g_{\bar{I}11}$,
$C_{\bar{I}\bar{J}\bar{K}}$ are odd. We note that the above boundary
actions contain, in addition to the Yang-Mills terms, $\tr R^2$ terms which
were not part of the original theory
derived in~\cite{hw2}. It was argued in ref.~\cite{low1} that these terms
are required by supersymmetry, since they pair with the $R^2$ terms in the
Bianchi identity~\eqref{Bianchi} in analogy to the weakly coupled case.
The existence of these terms will be of some importance in the following.

One way to view this theory is to draw an analogy between the orbifold
planes and D-branes in Type II theories. A collection of D$p$-branes
is described by a $U(N)$ gauge theory. The D$p$-brane charge is
measured by $\tr{\bf 1}=N$, while exciting a D$(p-2)$-brane charge
corresponds to having a non-trivial $\tr F$, and a D$(p-4)$-brane
charge corresponds to non-trivial $\tr F\w F$ and
so on~\cite{li1}. Similarly, if the original D-branes are on a curved
manifold then there is also an induced charge for lower-dimensional
branes given by $\tr R\w R$ and higher even powers~\cite{ghm}.
Applying this picture to our situation,  the r\^ole of the $U(N)$ gauge
field on the $D$-brane worldvolume is here played by the $E_8$ gauge fields
on the orbifold planes. The correction to the Bianchi
identity then has the interpretation of exciting an M5-brane charge in
the orbifold plane. In ref.~\cite{llo} this picture has been made explicit by
constructing a gauge five-brane in this theory.

We would now like to discuss solutions of the above theory which preserve
four of the 32 supercharges leading, upon compactification, to four
dimensional $N=1$ supergravities. This task is significantly complicated by
the fact that the sources in the Bianchi identity~\eqref{Bianchi} are
located on the orbifold planes with the gravitational part distributed
equally between the two planes. While the standard embedding of the
spin connection into the gauge connection
\begin{equation}
 \tr F^{(1)}\w F^{(1)} = \tr R\w R
\label{condition}
\end{equation}
leads to vanishing source terms in the weakly coupled heterotic string
Bianchi identity (which, in turn, allows one to set the antisymmetric tensor
gauge field to zero), in the present case, one is left with non-zero sources
$\pm \frac{1}{2} \tr R\w R$ on the two hyperplanes. As a result, the 
antisymmetric tensor
field $G$ and, hence, the second term in the gravitino supersymmetry 
variation
\begin{equation}
 \d\J_I = D_I\eta +\frac{\sqrt{2}}{288}\left(\G_{IJKLM}-8g_{IJ}\G_{KLM}
          \right)G^{JKLM}\eta + \; \cdots \; ,
\label{susy}
\end{equation}
do not vanish. Thus, straightforwardly compactifying on a Calabi-Yau manifold
no longer provides a solution to the Killing spinor equation $\d\J_I =0$.
The problem can, however, be treated perturbatively in powers
of the eleven-dimensional 
Newton constant $\k$. To lowest order, one can start with a manifold
$M_4\times S^1/Z_2\times X$ where $X$ is a Calabi-Yau three-fold and
$M_4$ is four-dimensional Minkowski space. This manifold has an
$x^{11}$-independent (and hence chiral) Killing spinor $\eta$ which
corresponds to four preserved supercharges. Then, one can determine the 
first order corrections to this background and the spinor $\eta$ so that
the gravitino variation vanishes to order $\k^{2/3}$.

The existence of such a ``deformed'' background solution to order
$\k^{2/3}$  has been demonstrated in ref.~\cite{w}. To see its explicit
form, let us start with the zeroth order metric
\begin{equation}
 ds_{11}^2 = \eta_{\m\n}dx^\m dx^\n+R_0^2(dx^{11})^2+V_0^{1/3}\O_{AB}dx^Adx^B
             \; ,
\end{equation}
where $\O_{AB}$ is a Calabi-Yau metric with K\"ahler form
$\o_{a\bar{b}}=i\O_{a\bar{b}}$. (Here $a$ and $\bar{b}$ are holomorphic
and anti-holomorphic indices.) To keep track of the scaling properties of 
the solution, we have introduced moduli $V_0$ and $R_0$ for the Calabi-Yau
volume and the orbifold radius, respectively. It was shown in~\cite{w}
that, to order $\k^{2/3}$, the metric can be written in the form
\begin{equation}
 ds_{11}^2 = (1+\hat{b})\eta_{\m\n}dx^\m dx^\n+R_0^2(1+\hat{\g})(dx^{11})^2
      +V_0^{1/3}(\O_{AB}+h_{AB})dx^Adx^B \label{metric}
\end{equation}
where the functions $\hat{b}$, $\hat{\g}$ and $h_{AB}$ depend on $x^{11}$
and the Calabi-Yau coordinates. Furthermore, as we have discussed, $G_{ABCD}$
and $G_{ABC11}$ receive a contribution of order $\k^{2/3}$ from the Bianchi
identity source terms. To get the general explicit form of the corrections,
one has to solve the relations given in ref.~\cite{w}. This can be done by
dualizing the antisymmetric tensor field and using a harmonic expansion on the
Calabi-Yau space~\cite{low1}. 

Here, we quote those results simplified in two
essential ways. First, we drop all terms corresponding to non-zero
eigenvalue harmonics on the Calabi-Yau space. These terms will be of no
relevance to the low energy theory, since they correspond to heavy
Calabi-Yau modes which decouple at this order. Second, we write only
the one massless term that is related to the Calabi-Yau breathing
mode. This will be sufficient for all applications dealing only with the
universal moduli. Given these simplifications, the corrections are
explicitly 
\begin{align}
 \hat{b} &= -\frac{\sqrt{2}R_0}{3V_0^{2/3}}\a_{0}\, (|x^{11}|-\p\r /2)
    \label{corra} \\
 \hat{\g} &= \frac{2\sqrt{2}R_0}{3V_0^{2/3}}\a_{0}\, (|x^{11}|-\p\r /2)
    \label{corrb} \\
 h_{AB} &= \frac{\sqrt{2}R_0}{3V_0^{2/3}}\a_{0}\, (|x^{11}|-\p\r
    /2)\O_{AB}
    \label{corrc} \\
 G_{ABCD} &= \frac{1}{6}\a_{0}\,{\e_{ABCD}}^{EF}\,\o_{EF}\,\e (x^{11})
    \label{corrd} \\
 G_{ABC11} &= 0 
    \label{corre} 
\end{align}
with
\begin{equation}
\begin{aligned}
 \a_{0} &= -\frac{1}{8\sqrt{2}\p v}\left(\frac{\k}{4\p}\right)^{2/3}\int_X
      \o\wedge\tr R^{(\O )}\wedge R^{(\O )}\; , \\
 v &= \int_X\sqrt{\O}\; .\label{alpha}
\end{aligned}
\end{equation}
Here $\e (x^{11})$ is the step function which is $+1$ ($-1$) for $x^{11}$
positive (negative).
Note that, by dropping the massive modes, these expressions take
a very simple form representing a linear increase of the corrections along
the orbifold. Even more significantly, and unlike the exact solution
including the heavy modes, the above approximation leads to a corrected metric
$\O_{AB}+h_{AB}$ that is still of Calabi-Yau type  at each point on the 
$S^1/Z_2$ orbifold. The Calabi-Yau volume (and, if all moduli are included,
also its shape), however, is continuously changing across the orbifold. More
generally, one can think of the internal part of the corrected metric as a
curve in the Calabi-Yau moduli space.

Returning to the D-brane perspective, one can view the above configuration as
the linearized solution for a collection of five-branes embedded in the
orbifold planes. The relation~\eqref{condition} fixes equal amounts of
five-brane charge, $\frac{1}{2}\tr R\w R$, on each orbifold fixed plane, where
the five-branes are confined to live. Since $\tr R\w R \in
H^{2,2}(X)$, we can associate a different five-brane charge for each
independent element of $H^{2,2}(X)$. The five-branes themselves are
associated with Poincar\'e dual cycles. Thus they span the non-compact
four-dimensional space together with a two-cycle in the Calabi-Yau space.
In particular, from the five-dimensional point of view, they are three-branes
localized on the orbifold planes. Witten's construction ensures that this
configuration of branes preserves one-eighth of the supersymmetry. Finally,
restricting to just the Calabi-Yau breathing modes corresponds to keeping only
the five-brane which spans the holomorphic two-cycle in the
Calabi-Yau three--fold defined by the K\"ahler form. 


\section*{The five-dimensional effective action}


Phenomenologically, there is a regime where the universe appears
five-dimensional. We would, therefore, like to derive an effective
theory in the space consisting of the usual four space-time dimensions
and the orbifold, based on the background solution discussed in the
previous section. As we have already mentioned, we will consider the
universal zero modes only; that is, the five-dimensional graviton
supermultiplet and the breathing mode of the Calabi-Yau space, along
with its superpartners. These form a hypermultiplet in five
dimensions. Furthermore, to keep the discussion as simple as possible,
we will not consider boundary gauge matter fields. This simple
framework suffices to illustrate our main ideas. The general case was
presented in~\cite{losw1}. 

Naively, one might attempt to perform the actual reduction directly on
the background given in eqs.~\eqref{metric} and \eqref{corra}--\eqref{corre}.
This would, however, lead to a complicated five-dimensional theory with
explicit $x^{11}$-dependence in the action. Moreover, this background
preserves only four supercharges whereas the minimal supergravity in
five dimensions ($N=1$) is invariant under twice this amount of supersymmetry.

A useful observation here is that, since we retain the dependence on the
orbifold coordinate, we can actually absorb the metric deformations in
\eqref{metric} and \eqref{corra}--\eqref{corre} into the
five-dimensional metric moduli. That 
is, the $x^{11}$-dependent scale factors $\hat{b}$ and $\hat{\g}$ of the
four-dimensional space and of the orbifold can be absorbed into the
five-dimensional (Einstein frame) metric $g_{\a\b}$ while,
analogously, the variation of the Calabi-Yau volume along the orbifold
encoded in $h_{AB}$ can be absorbed into a modulus
$V$.~\footnote{Note that we could not apply a similar method for a
reduction down to four dimensions, as all moduli fields would then be
$x^{11}$ independent. In this case, one should work with the
background in the form \eqref{metric}, \eqref{corra}--\eqref{corre} as
done in ref.~\cite{low1}.} More precisely, we can perform the Kaluza-Klein
reduction on the metric 
\begin{equation}
 ds_{11}^2 = V^{-2/3}g_{\a\b}dx^\a dx^\b +V^{1/3}\O_{AB}dx^Adx^B\; .
 \label{metric1}
\end{equation}
This rewriting suggests a change of perspective: rather than reducing on the
Witten vacuum, we can try to find an effective five-dimensional theory
where we recover the Witten vacuum as a particular solution.

We see that, since we have absorbed the deformation into the
moduli, the background corresponding to the metric~\eqref{metric1}
preserves eight supercharges, the appropriate number for a reduction down to
five dimensions. It might appear that we are
simply performing a standard reduction of eleven-dimensional supergravity
on a Calabi-Yau space to five dimensions; for example, in the way described in
ref.~\cite{CYred}. If this were the case, then it would be hard to understand
how the resulting five-dimensional theory could encode any information about
the deformed Calabi-Yau background. There are, however, two important
ingredients that we have not yet included. One is obviously the existence of
the boundary theories. We will return to this point shortly. First, however,
let us explain a somewhat unconventional addition to the bulk theory that
must be included.

Although we could absorb all metric corrections into the five-dimensional
metric moduli, the same is not true for the 4-form
field. Specifically, for the nonvanishing component $G_{ABCD}$ in
eq.~\eqref{corrd} there is no corresponding zero mode
field~\footnote{This can be seen from the mixed part of the Bianchi
identity $\partial_\a G_{ABCD}=0$ which shows that the constant $\a_{0}$
in eqs.~\eqref{corra}--\eqref{corre} cannot be promoted as stands to 
a five-dimensional field. It is possible to dualize in five dimensions
so the constant $\a_{0}$ is promoted to a five-form field, but we will not
pursue this formulation here.}.
Therefore, in the reduction, we should take this part of $G$ explicitly
into account. In the terminology of ref.~\cite{gsw}, such an antisymmetric
tensor field configuration is called a ``non-zero mode''. More generally, a
non-zero mode is a background antisymmetric tensor field that solves the
equations of motion but, unlike antisymmetric tensor field moduli, has
nonvanishing field strength. Such configurations, for a $p$-form field
strength, can be identified with the cohomology group $H^p(M)$ of the manifold
$M$ and, in particular, exist if this cohomology group is nontrivial. In the
case under consideration, the relevant cohomology group is $H^4(X)$ which
is nontrivial for a Calabi-Yau manifold $X$ since $h^{2,2}=h^{1,1}\geq 1$.
Again, the form of $G_{ABCD}$ in eq.~\eqref{corrd} is somewhat special,
reflecting the fact that we are concentrating here on the universal moduli. In
the general case, $G_{ABCD}$ would be a linear combination of all
harmonic $(2,2)$-forms.

The complete configuration for the antisymmetric tensor field that we use in
the reduction is given by
\begin{equation}
\begin{aligned}
 C_{\a\b\g} &, \quad 
   G_{\a\b\g\d}=24\,\partial_{[\a}C_{\b\g\d]} \\
 C_{\a AB} = \frac{1}{6}\cA_\a\,\o_{AB} &, \quad 
   G_{\a\b AB}=\cF_{\a\b}\,\o_{AB}=2\partial_{[\a}\cA_{\b]}\,\o_{AB}\; , \\
 C_{ABC} = \frac{1}{6}\x\,\o_{ABC} &,\quad 
   G_{\a ABC}=\partial_\a\x\,\o_{ABC} 
\end{aligned}
\label{Gmod}
\end{equation}
and the non-zero mode is
\begin{equation}
 G_{ABCD} = \frac{\a_{0}}{6}{\e_{ABCD}}^{EF}\,\o_{EF}\,\epsilon (x^{11})\; ,
            \label{nonzero}
\end{equation}
where $\a_{0}$ was defined in eq.~\eqref{alpha}. Here, $\o_{ABC}$ is the
harmonic $(3,0)$ form on the Calabi-Yau space and $\x$ is the corresponding
(complex) scalar zero mode. In addition, we have a five-dimensional vector
field $\cA_\a$ and 3-form $C_{\a\b\g}$, which can be
dualized to a scalar $\s$. The total bulk field content of the
five-dimensional theory is then given by the gravity multiplet
$(g_{\a\b},\cA_\a ,\psi^i_\a )$ together with the universal hypermultiplet
$(V,\s ,\x ,\bar{\x},\z^i)$ where $\psi_\a^i$ and
$\z^i$ are the gravitini and the hypermultiplet fermions respectively and
$i=1,2$. From their relations to the eleven-dimensional fields, it is easy 
to see that $g_{\m\n}$, $g_{11,11}$, $\cA_{11}$, $\s$ must be even under the
$Z_2$ action whereas $g_{\m 11}$, $\cA_\m$, $\x$ must be odd.

Examples of compactifications with non-zero modes in pure eleven-dimensional
supergravity on various manifolds including Calabi-Yau three-folds have
been studied in ref.~\cite{llp}. There is, however, one important way in
which our non-zero mode differs from other non-zero
modes in pure eleven-dimensional supergravity. Whereas the latter may be viewed
as an optional feature of generalized Kaluza-Klein reduction, the non-zero
mode in Ho\v{r}ava-Witten theory that we have identified cannot be turned
off. This can be seen from the fact that the constant $\a_{0}$ in
expression~\eqref{nonzero} cannot be set to zero. This is unlike the case in pure
eleven-dimensional supergravity where it would be arbitrary, since it is 
fixed by eq.~\eqref{alpha} in terms of Calabi-Yau data. This fact is, of 
course, intimately related to the existence of the boundary source terms, 
particularly in the Bianchi identity~\eqref{Bianchi}. As we will see, 
keeping the non-zero mode in the derivation of the five-dimensional action 
is crucial to finding a solution of this theory that corresponds to the 
deformed Calabi-Yau space discussed in the previous section.

Let us now turn to a discussion of the boundary theories. In the
five-dimensional space $M_5$ of the reduced theory, the orbifold fixed
planes constitute four-dimensional hypersurfaces which we denote by
$M_4^{(i)}$, $i=1,2$. Clearly, since we have used the standard embedding,
there will be an $E_6$ gauge field $A_\m^{(1)}$accompanied by gauginos and
gauge matter fields on the orbifold plane $M_4^{(1)}$. For simplicity,
we will set these gauge matter fields to zero in the following. The field
content of the orbifold plane $M_4^{(2)}$ consists of an $E_8$ gauge field
$A_\m^{(2)}$ and the corresponding gauginos. In addition, there is another
important boundary effect which results from the non-zero internal gauge
field and gravity curvatures. More precisely, note that
\begin{equation}
\begin{aligned}
 \int_X\sqrt{\O}\, \tr F_{AB}^{(1)}F^{(1)AB} 
    &= \int_X \sqrt{\O}\,\tr R_{AB}R^{AB} 
     = -16\sqrt{2}\p v\left(\frac{4\p}{\k} \right)^{2/3}\a_{0}\; , \\
 F_{AB}^{(2)} &= 0\; .
\end{aligned}
\label{intcurv}
\end{equation}
In view of the boundary actions~\eqref{SYM}, it follows that we will retain
cosmological type terms with opposite signs on the two boundaries.
Note that the size of those terms is set by the same constant $\a_{0}$,
given by eq.~\eqref{alpha}, which determines the magnitude of the non-zero
mode. The boundary cosmological terms are another important ingredient in
reproducing the eleven-dimensional background as a solution of the
five-dimensional theory.

We can now compute the five-dimensional effective action of
Ho\v rava-Witten theory. Using the field
configuration~\eqref{metric1}--\eqref{intcurv} we find from the
action~\eqref{action}--\eqref{SYM} that
\begin{equation}
 S_5 = S_{\rm grav}+S_{\rm hyper}+S_{\rm bound}\label{S5}
\end{equation}
where
\begin{align}
 S_{\rm grav} =& -\frac{1}{2\k_5^2}\int_{M_5}\sqrt{-g}\left[
     R + \frac{3}{2}\cF_{\a\b}\cF^{\a\b}
     + \frac{1}{\sqrt{2}}
        \e^{\a\b\g\d\e}\cA_\a\cF_{\b\g}\cF_{\d\e}\right] 
    \label{actpartsa} \\
 S_{\rm hyper} =& -\frac{1}{2\k_5^2}\int_{M_5}\sqrt{-g}\left[
                   \frac{1}{2V^2}\partial_\a V\partial^\a V
                   +\frac{2}{V}\partial_\a\x\partial^\a\bar{\x}
                   +\frac{V^2}{24}G_{\a\b\g\d}G^{\a\b\g\d}
                   \right.\nn \\ & \qquad {}
                    +\frac{\sqrt{2}}{24}\e^{\a\b\g\d\e}G_{\a\b\g\d}
                   \left(i(\x\partial_\e\bar{\x}-\bar{\x}\partial_\e\x )+
                   2\a_{0}\e(x^{11})\cA_\e\right)+\frac{1}{3V^2}\a_{0}^2\bigg]
    \label{actpartsb} \\
 S_{\rm bound} =& \frac{\sqrt{2}}{\k_5^2}\int_{M_4^{(1)}}\sqrt{-g}
                   \, V^{-1}\a_{0}
                  - \frac{\sqrt{2}}{\k_5^2}\int_{M_4^{(2)}}\sqrt{-g}\,
                     V^{-1}\a_{0} \nn \\ & \qquad
                  -\frac{1}{16\p\a_{\rm GUT}}
                   \sum_{i=1}^2\int_{M_4^{(i)}}\sqrt{-g}\, V\tr
                   {F_{\m\n}^{(i)}}^2 \; .
    \label{actpartsc}
\end{align}
In this expression, we have now dropped higher-derivative terms. The
four-form field strength $G_{\a\b\g\d}$ is subject to the Bianchi identity
\begin{equation}
 (dG)_{11\m\n\r\s} = 
     -\frac{\k_5^2}{4\sqrt{2}\pi\a_{\rm GUT}}\left\{ 
       J^{(1)} \d (x^{11}) + J^{(2)} \d (x^{11}-\pi\r )
       \right\}_{\m\n\r\s} \label{Bianchi5}
\end{equation}       
which follows directly from the eleven-dimensional Bianchi
identity~\eqref{Bianchi}. The currents $J^{(i)}$ have been defined in
eq.~\eqref{J}. The five-dimensional Newton constant $\k_5$ and the
Yang-Mills coupling $\a_{\rm GUT}$ are expressed in terms of
eleven-dimensional quantities as~\footnote{The following relations are given
for the normalization of the eleven-dimensional action as in 
eq.~\eqref{action}. If instead the normalization of~\cite{conrad} is used, 
the expression for $\a_{\rm GUT}$ gets rescaled to 
$a_{\rm GUT}=2^{1/3}\left(\k^2/2v\right)\left(4\p/\k\right)^{2/3}$ 
Otherwise the action and Bianchi identities are unchanged, except that in 
the expression~\eqref{intcurv} for $\alpha$ the RHS is multiplied 
by $2^{1/3}$.}
\begin{equation}
 \k_5^2=\frac{\k^2}{v}\; ,\qquad \a_{\rm GUT} = \frac{\k^2}{2v}\left(
   \frac{4\p}{\k}\right)^{2/3}\; .
\end{equation}
We have checked the consistency of the truncation which leads
to the above action by an explicit reduction of the eleven-dimensional
equations of motion to five dimensions. Note that the potential terms in the
bulk and on the boundaries arise precisely from the inclusion of the
non-zero mode and the gauge and gravity field strengths, respectively.
Since we have compactified on a Calabi-Yau space, we expect
the bulk part of the above action to have eight preserved supercharges
and, therefore, to correspond to minimal $N=1$ supergravity in five
dimensions. Accordingly, let us compare the result
\eqref{actpartsa}--\eqref{actpartsc} to the known $N=1$
supergravity-matter theories in five
dimensions~\cite{GST1,GST2,Sierra}.

In these theories, the scalar fields in the universal hypermultiplet 
parameterize a quaternionic manifold with coset structure
$\cM_Q=SU(2,1)/SU(2)\times U(1)$. Hence, to compare our action to these
we should dualize the three-form $C_{\a\b\g}$ to a scalar field $\s$ by
setting (in the bulk)
\begin{equation}
 G_{\a\b\g\d} = \frac{1}{\sqrt{2}V^2}\e_{\a\b\g\d\e}\left(\partial^\e\s
                -i(\x\partial^\e\bar{\x}-\bar{\x}\partial^\e\x )-2
                \a_{0}\e(x^{11})\cA^\e\right)\; .
\end{equation}
Then the hypermultiplet part of the action \eqref{actpartsb} can be written as
\begin{equation}
 S_{\rm hyper} = -\frac{v}{2\k^2}\int_{M_5}\sqrt{-g}\left[ h_{uv}\nabla_\a
                 q^u\nabla^\a q^v +\frac{1}{3}V^{-2}\a_{0}^2\right]
 \label{hyper}
\end{equation}
where $q^u=(V,\s ,\x ,\bar{\x})$. The covariant derivative $\nabla_\a$ is
defined as $\nabla_\a q^u= \partial_\a q^u+\a_{0}\e(x^{11})\cA_\a k^u$ with
$k^u=(0,-2,0,0)$. The sigma model metric 
$h_{uv} = \partial_u\partial_vK_Q$
can be computed from the K\"ahler potential
\begin{equation}
   K_Q = -\ln (S+\bar{S}-2C\bar{C})\; , \quad
   S = V+\x\bar{\x}+i\s\; , \quad 
   C = \x\; .
\end{equation}
Consequently, the hypermultiplet scalars $q^u$ parameterize a K\"ahler
manifold with metric $h_{uv}$. It can be demonstrated that $k^u$ is a Killing vector
on this manifold. Using the expressions given in ref.~\cite{strom}, one can
show that this manifold is quaternionic with coset structure $\cM_Q$.
Hence, the terms in eq.~\eqref{hyper} that are independent of $\a_{0}$ describe the
known form of the universal hypermultiplet action. How do we interpret
the extra terms in the hypermultiplet action depending on
$\a_{0}$? A hint is provided by the fact that one of these $\a_{0}$-dependent terms
modifies the flat derivative in the kinetic energy to
a generalized derivative $\nabla_\a$. This is exactly the
combination that we would need if one wanted to gauge the $U(1)$ symmetry on
$\cM_Q$ corresponding to the Killing vector $k^u$, using the gauge field
$\cA_\a$ in the gravity supermultiplet. In fact,
investigation of the other terms in the action, including the
fermions, shows that the resulting five-dimensional theory is
precisely a gauged form of supergravity. Not only is a $U(1)$ isometry
of $\cM_Q$ gauged, but at the same time a $U(1)$ subgroup
of the $SU(2)$ automorphism group is also gauged. 

What about the remaining $\a_{0}$-dependent potential term in the
hypermultiplet action? From $d=4$, $N=2$  theories, we are used to the
idea that gauging a symmetry of the quaternionic manifold describing 
hypermultiplets generically introduces potential terms into the
action when supersymmetry is preserved (see for
instance~\cite{andetal}). Such potential terms can be thought of as
the generalization of pure Fayet-Iliopoulos (FI) terms. This is precisely what
happens in our theory as well, with the gauging of the  $U(1)$ subgroup
inducing the $\a_{0}$-dependent potential term in ~\eqref{hyper}.
The general gauged action was discussed in detail in~\cite{losw1}.
Certain pure FI terms were previously considered in~\cite{GST2}, but, to our
knowledge, such a theory with general gauging has not been constructed
previously in five dimensions. 

The phenomenon that the inclusion of non-zero modes leads
to gauged supergravity theories has already been observed in Type II
Calabi-Yau compactifications~\cite{sp,michelson}, while the
observation that the vacua of gauged theories correspond to
dimensional reduction with non-trivial form-fields has a long
history. Recent results relating to intersecting branes 
are described in~\cite{ct}. From the form of the Killing vector,
we see that it is only the scalar field $\s$, dual to the four-form
$G_{\a\b\g\d}$, which is charged under the $U(1)$ symmetry. Its charge
is fixed by $\a_{0}$. We note that this charge is quantized since,
suitably normalized, $\tr R\w R$ is an element of $H^{2,2}(X,{\bf
Z})$. In the brane description of the theory, this is a reflection of
the fact that the five-brane charge is quantized.

To analyze the supersymmetry properties of the solutions shortly to be
discussed, we need the supersymmetry variations of the fermions
associated with the theory~\eqref{S5}. They can be obtained either by
a reduction of the eleven-dimensional gravitino variation~\eqref{susy} or
by generalizing the known five-dimensional transformations~\cite{GST1,Sierra}
by matching onto gauged four-dimensional $N=2$ theories. It is
sufficient for our purposes to keep the bosonic terms only. Both approaches
lead to
\begin{equation}
\begin{aligned}
 \d \psi_\a^i =& D_\a\e^i 
     + \frac{\sqrt{2}i}{8}
          \left({\g_\a}^{\b\g}-4\d_\a^\b\g^\g\right)\cF_{\b\g}\e^i
     - \frac{1}{2}V^{-1/2}\left(
        \partial_\a\x\,{(\t_1-i\t_2)^i}_j
        - \partial_\a{\bar\x}\,{(\t_1+i\t_2)^i}_j \right) \e^j
     \\ &
     - \frac{\sqrt{2}i}{96}V{\e_\a}^{\b\g\d\e}G_{\b\g\d\e}{(\t_3)^i}_j\e^j
     - \frac{\sqrt{2}}{12}\a_{0} V^{-1}\e (x^{11})\g_\a{(\t_3)^i}_j\e^j 
     \\
 \d\z^i =& \frac{\sqrt{2}}{48}V\e^{\a\b\g\d\e}G_{\a\b\g\d}\g_\e\e^i
     - \frac{i}{2}V^{-1/2}\g^\a\left(
        \partial_\a\x\,{(\t_1-i\t_2)^i}_j
        + \partial_\a{\bar\x}\,{(\t_1+i\t_2)^i}_j \right) \e^j
     \\ &
     + \frac{i}{2}V^{-1}\g_\b\partial^\b V\e^i
     - \frac{i}{\sqrt{2}}\a_{0} V^{-1}\e (x^{11}){(\t_3)^i}_j\e^j 
\end{aligned}
\label{susy5} 
\end{equation}
where $\t_i$ are the Pauli spin matrices. 

In summary, we see that the relevant five-dimensional effective theory
for the reduction of Ho\v{r}ava-Witten theory is a gauged $N=1$ supergravity
theory
with bulk and boundary potentials. While we have calculated the theory only to
order $\k^{2/3}$, one would expect that M-theory corrections can be
described in the same type of theory. For this reason, it is very
desirable to construct the most general gauged five-dimensional $N=1$
supergravity theory coupled to general $N=1$ four-dimensional boundary
theories with vector and chiral multiplets. Part of this program 
was carried out in ref.~\cite{losw1}. In the context of global
supersymmetry, such boundary theories in five dimensions have been studied
in ref.~\cite{sharpe}.  


\section*{Lecture 2:}


In this second lecture, we discuss the static vacuum solution of the effective
five--dimensional Ho\v rava-Witten theory. We show that it is given by a pair
of BPS domain walls, each located at one orbifold boundary plane. At leading
non-trivial order, this domain wall solution is identical to the
``deformations'' of the eleven--dimensional background spacetime found by
Witten. The reduction of the five--dimensional theory on this BPS domain
wall then produces, at low energy, the four--dimensional $N=1$ supersymmetric
effective heterotic $M$-theory. We then present, in detail, the simplest
time-dependent cosmological vacuum solution associated with this domain wall
background. We discuss various aspects of its cosmology and show that it has
at least one branch that corresponds to a radiation--like expanding
three--dimensional universe and a contracting orbifold.

\section*{The domain-wall solution}


Let us recapitulate what we have done so far. To arrive at a simple form
for the five dimensional effective action, we have absorbed the
deformation of the Calabi-Yau background metric into the 
five-dimensional moduli. Effectively, we could then carry out the reduction
on a Calabi-Yau space but had to explicitly keep the antisymmetric tensor
part of the background as a non-zero mode in the reduction.
As a consequence, although Witten's original background preserved only
four supercharges, the effective bulk theory has twice that number
of preserved supercharges, corresponding to minimal $N=1$ supergravity in five
dimensions. For consistency, we should now be able to find the
deformations of the Calabi-Yau background as solutions of
the effective five-dimensional theory. These solutions should break half
the supersymmetry of the five-dimensional bulk theory and preserve Poincar\'e
invariance in four dimensions. Hence, we expect there to be a three-brane
domain wall in five dimensions with a worldvolume lying in the four
uncompactified directions. This domain wall can be viewed as the ``vacuum'' of
the five-dimensional theory, in the sense that it provides the appropriate
background for a reduction to the $d=4$, $N=1$ effective theory.

This expectation is made stronger if we recall the brane picture
of Witten's background. We argued that this could be
described by five-branes with equal amounts of five-brane
charge living on the orbifold planes. From the five-dimensional perspective,
the five-branes appear as three-branes living on the orbifold fixed planes.
Thus, in five dimensions, Witten's background must correspond to a pair of
parallel three-branes. 

We notice that the theory~\eqref{S5} has all of the prerequisites necessary for
such a three-brane solution to exist. Generally, in order to have a
$(D-2)$-brane in a $D$-dimensional theory, one needs to have a $(D-1)$-form
field or, equivalently, a cosmological constant. This is familiar from the
eight-brane~\cite{8brane} in the massive type IIA supergravity in ten
dimensions~\cite{romans}, and has been systematically studied for theories in
arbitrary dimension obtained by generalized (Scherk-Schwarz) dimensional
reduction~\cite{dom}. In our case, this cosmological term is provided by
the bulk potential term in the action~\eqref{S5}. From the
viewpoint of the bulk theory, we could have multi three-brane solutions with
an arbitrary number of parallel branes located at various places in the
$x^{11}$ direction. As is well known, however, elementary brane solutions have
singularities at the location of the branes, needing to be supported by
source terms. The natural candidates for those source terms, in our case, are
the boundary actions. Given the anomaly-cancelation requirements,
this restricts the possible solutions to those representing a pair of
parallel three-branes corresponding to the orbifold planes.

{}From the above discussion, it is clear that in order to find a three-brane
solution, we should start with the Ansatz
\bea
 ds_5^2 &=& a(y)^2dx^\m dx^\n\eta_{\m\n}+b(y)^2dy^2  \\
 V &=& V(y)\nn
\label{final2}
\eea
where $a$ and $b$ are functions of $y=x^{11}$ and all other fields vanish.
The general solution for this Ansatz, satisfying the equations of motion
derived from action~\eqref{S5}, is given by
\bea
 a &=&a_0H^{1/2}\nn \\
 b &=& b_0H^2\qquad\qquad H=\frac{\sqrt{2}}{3}\a_{0}|y|+c_0 \label{sol}\\
 V &=&b_0H^3 \nn
\label{final3}
\eea
where $a_0$, $b_0$ and $c_0$ are constants. We note that the boundary
source terms have fixed the form of the harmonic function $H$ in the
above solution. Without specific information about the sources, the function
$H$ would generically be glued together from 
an arbitrary number of linear pieces with
slopes $\pm\frac{\sqrt{2}}{3}$$\a_{0}$. The edges of each piece would then indicate
the location of the source terms. The necessity of matching the boundary
sources at $y=0$ and $\p\r$, however, has forced us to consider only two such
linear pieces, namely $y\in [0,\p\r ]$ and $y\in [-\p\r ,0]$. These pieces are
glued together at $y=0$ and $\p\r$
(recall here that we have identified $\p\r$ and $-\p\r$). 
To see this explicitly, let us consider one of the equations of motion; 
specifically, the equation derived from the variation of
$g_{\mu\nu}$. For the Ansatz in~\eqref{final2}, this is given by
\begin{equation}
 \frac{a''}{a}+\frac{{a'}^{2}}{a^{2}}-\frac{a'}{a}
  \frac{b'}{b}+\frac{1}{12}\frac{{V'}^{2}}{V^{2}}+
  \frac{\alpha_{0}^{2}}{18}\frac{b^{2}}{V^{2}} = \frac{\sqrt{2}\alpha_{0}}{3}
 \frac{b}{V}\left(\delta(y)-\delta(y-\pi\rho)\right)
\label{burt3}
\end{equation}
where the prime denotes differentiation with respect to $y$. The term 
involving the delta functions arises from the stress energy on the boundary 
planes.
Inserting the solution~\eqref{final3} in this equation, we have 
\bea
\partial_y^2H &=& \frac{2\sqrt{2}}{3}\a_{0}(\d (y)-\d (y-\p\r )) 
\eea
which shows
that the solution represents two parallel three-branes located at the
orbifold planes. 

We stress that this solution solves the five-dimensional
theory~\eqref{S5} exactly, whereas the original deformed Calabi-Yau
solution was only an approximation to order $\k^{2/3}$.
It is straightforward to show that the linearized version of~\eqref{sol},
that is, the expansion to first order in $\a_{0} =O(\k^{2/3})$, coincides with
Witten's solution~\eqref{metric}, \eqref{corra}--\eqref{corre} upon
appropriate matching of the integration constants. Hence, we have
found an exact generalization, good to all orders in $\k$, of the
linearized eleven-dimensional solution. 

Of course, we still have to check that our solution preserves
half of the supersymmetries. When $g_{\a\b}$ and $V$ are the only non-zero
fields, the supersymmetry transformations~\eqref{susy5} simplify to
\bea
 \d\psi_\a^i &=&  D_\a\e^i -\frac{\sqrt{2}}{12}\a_{0}\,\e (y)V^{-1}\g_\a\,
                  {(\t_3)^i}_j\e^j\\
 \d\z^i &=& \frac{i}{2}V^{-1}\g_\b\partial^\b V\e^i-\frac{i}{\sqrt{2}}\a_{0}\,
            \e (y) V^{-1}\, {(\t_3)^i}_j\e^j\nn\; .
\label{eq:final1}
\eea
The Killing spinor equations $\d\psi_\a^i =0$, $\d\z^i=0$ are satisfied
for the solution~\eqref{sol} if we require that the spinor $\e^i$ is
given by
\begin{equation}
 \e^i = H^{1/4}\e^i_0\; ,\quad \g_{11}\e^i_0 = (\t_3)^i_j\e^j_0
\end{equation}
where $\e^i_0$ is a constant symplectic Majorana spinor. This shows that
we have indeed found a BPS solution preserving four of the eight bulk
supercharges.

Let us discuss the meaning of this solution in some detail. First, we notice
that it fits into the general scheme of domain wall solutions in various
dimensions~\footnote{In the notation of ref.~\cite{dom}, it corresponds
to choosing $D=5$, $\Delta = 4/3$ and $a(5)=2$.}. It is, however, a new
solution to the gauged supergravity action~\eqref{S5} in five dimensions
which has not been constructed previously. In addition, its source terms
are naturally provided by the boundary actions resulting from
Ho\v rava-Witten theory. Most importantly, it constitutes the fundamental
vacuum  solution of a phenomenologically relevant theory. The two parallel
three-branes of the solution, separated by the bulk, are oriented in
the four uncompactified space-time dimensions, and carry the physical
low-energy gauge and matter fields. Therefore, from the
low-energy point of view where the orbifold is not resolved the 
three-brane worldvolume is identified with four-dimensional space-time.
In this sense, the Universe lives on the worldvolume of a three-brane.

Although we have found an exact solution to the (lowest order) low energy
theory, thereby improving previous results, it is not clear whether the
solution will be exact in the full theory. Strominger~\cite{strom} has
argued that the
all-loop corrections (corresponding to corrections to the effective
action proportional to powers of $\k^{4/3}/V$, in our notation) to the
quaternionic metric of the universal hypermultiplet can be actually
absorbed into a shift of $V$, so that the metric is unchanged. This
implies that our solution would be unaffected by such corrections. On the
other hand, we have no general argument why the solution should be
protected against corrections from higher derivative terms. 

In any case, we believe that pursuing the construction of
five-dimensional gauged supergravities with boundaries, and the analysis
of their soliton structure, in the way indicated in this paper
will provide important insights into early universe cosmology
as well as low energy particle phenomenology.

It is the purpose of the remainder of these lectures to put this picture into
the context of cosmology. Consequently,
we are looking for time dependent solutions based on the static domain
wall which we have just presented.


\section*{The domain-wall cosmological solution}


In this section, we will present a cosmological solution related to the static
domain wall vacuum of the previous section. As discussed in
ref.~\cite{lowc1,lowc2}, a convenient way to find such a solution is to
use Ansatz~\eqref{final2} where the $y=x^{11}$ 
coordinate in the functions $a,b$ and $V$ is replaced by the time coordinate 
$\tau$. However, in Ho\v rava-Witten theory the boundary planes 
preclude this from being a solution of the equations of motion, since
it does not admit homogeneous solutions. To see this explicitly, let
us consider the $g_{00}$ equation of motion, where we replace $a(y)
\rightarrow \alpha(\tau), b(y) \rightarrow \beta(\tau)$ and $V
\rightarrow \gamma(\tau)$. We find that
\begin{equation}
 \frac{\dot{\alpha}^{2}}{\alpha^{2}}+\frac{\dot{\alpha}}{\alpha}
  \frac{\dot{\beta}}{\beta}-\frac{1}{12}\frac{\dot{\gamma}^{2}}{\gamma^{2}}
   -\frac{\alpha_{0}^{2}}{18}\frac{1}{\gamma^{2}}=
  -\frac{\sqrt{2}\alpha_{0}}{3}\frac{1}{\beta\gamma}\left(\delta(y)
  -\delta(y-\pi\rho)\right)\; ,
\label{burt4}
\end{equation}
where the dot denotes differentiation with respect to $\t$.
Again, the term containing the delta functions arises from the boundary planes.
It is clear that, because of the $y$-dependence introduced by the delta
functions, this equation has no globally defined solution. The structure of 
equation \eqref{burt4} suggests that a solution might be found if one were to 
let functions $a,b$ and $V$ depend on both $\tau$ and $y$ coordinates. This 
would be acceptable from the point of view of cosmology, since any such 
solution would be homogeneous and isotropic in the spatial coordinates $x^{m}$
where $m,n,r,\cdots = 1,2,3$. In fact, the previous Ansatz was too
homogeneous, being independent of the $y$ coordinate as well. Instead,
we are interested in solutions where the inhomogeneous vacuum domain
wall evolves in time. 

We now construct a cosmological 
solution where all functions depend on both $\tau$ and $y$. We start with the 
Ansatz
\bea
 ds_5^2 &=& -N(\tau,y)^{2}d\tau^{2}+a(\tau,y)^2dx^{m} dx^{n}\eta_{mn}+
            b(\tau,y)^2dy^2  \\
 V &=& V(\tau,y)\nn
 \label{burt5}
\eea
Note that we have introduced a separate function $N$ into the purely 
temporal part of the metric. 
This Ansatz leads to equations of motion that mix the $\tau$ and $y$
variables in a complicated non-linear way. In order to solve this system of 
equations, we will try to separate the two variables. That is, we let
\bea
 N(\tau,y)=n(\tau)a(y) \nn \\
 a(\tau,y)=\alpha(\tau)a(y) \nn\\
 b(\tau,y)=\beta(\tau)b(y) \label{burt6}\\
 V(\tau,y)=\gamma(\tau)V(y) \nn
\eea
There are two properties of this Ansatz that we wish to point out. 
The first is that for $n=\alpha=\beta=\gamma=1$ it becomes identical to
~\eqref{final2}. Secondly, we note that $n$ can be chosen to be any function 
by performing a redefinition of the $\tau$ variable. That is, we can 
think of $n$ as being subject to a gauge transformation. 
There is no a priori reason to believe that separation of variables will 
lead to a solution of the equations of motion derived from the
action~\eqref{S5}. However, as we now show, there is 
indeed such a solution. It is instructive to present one of the equations 
of motion. With the above Ansatz, the $g_{00}$ equation of motion is given
by~\footnote{From now on, we denote by $a$, $b$, $V$ the $y$-dependent
part of the Ansatz~\eqref{burt6}.}
\begin{multline}
\label{burt7}
 \frac{a^{2}}{b^{2}}\left(\frac{a''}{a}+\frac{{a'}^{2}}{a^{2}}-\frac{a'}{a}
  \frac{b'}{b}+\frac{1}{12}\frac{{V'}^{2}}{V^{2}}+\frac{\alpha_{0}^{2}}
   {18}\frac{b^{2}}{V^{2}}\frac{\beta^{2}}{\gamma^{2}}-\frac{\sqrt{2}}{3}
   \alpha_{0}\frac{b}{V}(\delta(y)-\delta(y-\pi\rho))\frac{\beta}{\gamma}
  \right) = \\
\frac{\beta^{2}}{n^{2}}\left(\frac{\dot{\alpha}^{2}}{\alpha^{2}}
 +\frac{\dot{\alpha}}{\alpha}\frac{\dot{\beta}}{\beta}-\frac{1}{12}
  \frac{\dot{\gamma}^{2}}{\gamma^{2}}\right)
\end{multline}
Note that if we set $n=\alpha=\beta=\gamma=1$ this equation becomes identical 
to \eqref{burt3}. Similarly, if we set $a=b=V=1$ and take the gauge $n=1$ 
this equation becomes the same as \eqref{burt4}. As is, the above 
equation does not separate. However, the obstruction to a separation of 
variables is the two terms proportional to $\alpha_{0}$. Note that both of 
these terms would be strictly functions of $y$ only if we demanded that
$\beta \propto \gamma$. Without loss of generality, one can take
\bea
\beta=\gamma\; .
\label{burt8}
\eea
We will, henceforth, assume this is the case. Note that this result is
already indicated by the structure of integration constants (moduli)
in the static domain wall solution~\eqref{final3}. With this condition, the
left hand side of equation \eqref{burt7} is purely $y$ dependent, whereas the 
right hand side is purely $\tau$ dependent. Both sides must now equal the 
same constant which, for simplicity, we take to be zero. The equation 
obtained by setting the left hand side to zero is identical to the pure 
$y$ equation \eqref{burt3}. The equation for the pure $\tau$ dependent 
functions is
\bea
 \frac{\dot{\alpha}^{2}}{\alpha^{2}}
 +\frac{\dot{\alpha}}{\alpha}\frac{\dot{\beta}}{\beta}-\frac{1}{12}
  \frac{\dot{\gamma}^{2}}{\gamma^{2}}=0
\label{burt9}
\eea
Hence, separation of variables can be achieved for the $g_{00}$ equation by 
demanding that \eqref{burt8} is true. What is more remarkable is that, 
subject to the constraint that $\beta=\gamma$, all the equations of motion 
separate. The pure $y$ equations are identical to those of the previous 
section and, hence, the domain wall solution~\eqref{final3} remains valid
as the $y$--dependent part of the solution. 

The full set of 
$\tau$ equations is found to be
\bea
 \frac{\dot{\alpha}^{2}}{\alpha^{2}}
 +\frac{\dot{\alpha}}{\alpha}\frac{\dot{\beta}}{\beta}-\frac{1}{12}
  \frac{\dot{\gamma}^{2}}{\gamma^{2}}=0
\label{burt10}
\eea
\bea
 2\frac{\ddot{\alpha}}{\alpha}-2\frac{\dot{\alpha}}{\alpha}\frac{\dot{n}}{n}
  +\frac{\ddot{\beta}}{\beta}-\frac{\dot{\beta}}{\beta}\frac{\dot{n}}{n}
   +\frac{\dot{\alpha}^{2}}{\alpha^{2}}+2\frac{\dot{\alpha}}{\alpha}
   \frac{\dot{\beta}}{\beta}+\frac{1}{4}\frac{\dot{\gamma}^{2}}
  {\gamma^{2}}=0
\label{burt11}
\eea
\bea
 \frac{\ddot{\alpha}}{\alpha}-\frac{\dot{\alpha}}{\alpha}\frac{\dot{n}}{n}
  +\frac{\dot{\alpha}^{2}}{\alpha^{2}}+\frac{1}{12}\frac{\dot{\gamma}^{2}}
  {\gamma^{2}}=0
\label{burt12}
\eea
\bea
 \frac{\ddot{\gamma}}{\gamma}+3\frac{\dot{\a}\dot{\g}}{\a\g}+
  \frac{\dot{\b}\dot{\g}}{\b\g}-\frac{\dot{\g}^2}{\g^2}
  -\frac{\dot{n}\dot{\g}}{n\g}=0
\label{burt13}
\eea
In these equations we have displayed $\beta$ and $\gamma$ independently, 
for reasons to become apparent shortly. Of course, one must solve these 
equations subject to the condition that $\beta=\gamma$.
As a first attempt to solve these equations, it is most convenient to choose 
a gauge for which
\bea 
n=\mbox{const}
\label{burt14}
\eea
so that $\t$ becomes proportional to the comoving time $t$, since
$dt=n(\t )d\t$. In such a gauge, the equations simplify considerably and
we obtain the solution
\bea
 \a &=&A |t-t_0|^p \nn\\
 \b &=&\g = B |t-t_0|^q \label{burt15}
\eea
where
\begin{equation}
 p=\frac{3}{11}(1\mp\frac{4}{3\sqrt{3}})\; ,\qquad
 q=\frac{2}{11}(1\pm2\sqrt{3})
\label{burt16}
\end{equation}
and $A$, $B$ and $t_0$ are arbitrary constants. 
We have therefore found a cosmological solution, based on the separation
Ansatz~\eqref{burt6}, with the $y$-dependent
part being identical to the domain wall solution~\eqref{final3} and the
scale factors $\a$, $\b$, $\g$ evolving according to the power
laws~\eqref{burt15}. This means that the shape of the domain wall pair stays
rigid while its size and the separation between the walls evolve in
time. Specifically, $\a$ measures
the size of the spatial domain wall worldvolume (the size of the
three-dimensional universe), while $\b$ specifies the separation of the
two walls (the size of the orbifold). Due to the separation constraint
$\g =\b$, the time evolution of the Calabi-Yau volume, specified by $\g$,
is always tracking the orbifold. From this point of view, we are
allowing two of the three moduli in~\eqref{final3}, namely $a_0$ and
$b_0$, to become time-dependent. Since these moduli multiply the harmonic
function $H$, it is then easy to see why a solution by separation of
variables was appropriate. 

To understand the structure of the above solution, it is useful to
rewrite its time dependent part in a more systematic way using
the formalism developed in ref.~\cite{lowc1,lowc2}. First, let us define
new functions $\hat{\alpha}, \hat{\beta}$ and $\hat{\g}$ by
\begin{equation}
 \alpha=e^{\hat{\alpha}}, \qquad  \beta=e^{\hat{\beta}}, \qquad
 \gamma=e^{6\hat{\g}}
\label{burt17}
\end{equation}
and introduce the vector notation
\begin{equation}
 \vec{\alpha}= (\a^i) =\left(
            \begin{array}{c}
            \hat{\alpha} \\
            \hat{\beta} \\    
            \hat{\g} \\
            \end{array}
            \right)\; ,\qquad
 \vec{d}= (d_i) = \left(
            \begin{array}{c}
            3 \\
            1 \\    
            0 \\
            \end{array}
            \right)\; .
 \label{burt18}
\end{equation} 
Note that the vector $\vec{d}$ specifies the dimensions of the various
subspaces, where the entry $d_1=3$ is the spatial worldvolume dimensions,
$d_2=1$ is the orbifold dimension and we insert $0$ for the dilaton.
On the ``moduli space'' spanned by $\vec{\alpha}$ we introduce the
metric
\bea
 G_{ij}&=&2(d_{i}\delta_{ij}-d_{i}d_{j}) \nn \\
 G_{in}&=&G_{ni}=0 \label{andre5}\\
 G_{nn}&=&36\; ,\nn
\eea
which in our case explicitly reads
\begin{equation}
G=-12 \left(
      \begin{array}{ccc}
      1  & \frac{1}{2} & 0  \\
      \frac{1}{2} & 0  & 0  \\
      0  & 0           & -3 \\
      \end{array}
      \right)\; .
\label{burt20}
\end{equation}
Furthermore, we define $E$ by
\begin{equation}
 E=\frac{e^{\vec{d}\cdot\vec{\alpha}}}{n}=
   \frac{e^{3\hat{\alpha}+\hat{\beta}}}{n}\; .
\label{burt19}
\end{equation}
The equations of motion \eqref{burt10}-\eqref{burt13} can then be
rewritten as
\begin{equation}
 \frac{1}{2}E\dot{\vec{\alpha}}^{T}G\dot{\vec{\alpha}}=0\; ,\qquad
 \frac{d}{d\t}\left(EG\dot{\vec{\alpha}}\right)=0\; .
\label{burt22}
\end{equation}
It is straightforward to show that if we choose a gauge $n=$ const, these
two equations exactly reproduce the solution given in~\eqref{burt15}
and~\eqref{burt16}. The importance of this reformulation of the
equations of motion lies, however, in the fact that we now get solutions
more easily by exploiting the gauge choice for $n$. For example, let us
now choose the gauge
\begin{equation}
 n = e^{\vec{d}\cdot\vec{\a}}\; .
\end{equation}
Note that in this gauge $E=1$. The reader can verify that this gauge choice
greatly simplifies solving the equations. The result is that
\bea
 \hat{\a}&=&6\hat{\g}=C\t +k_1 \nn \\
 \hat{\b}&=&(6\pm 4\sqrt{3})C\t +k_2
\eea
where $C$, $k_1$ and $k_2$ are arbitrary constants. Of course this solution
is completely equivalent to the previous one, eq.~\eqref{burt15}, but written
in a different gauge. We will exploit this gauge freedom to effect in the
last lecture.

To discuss cosmological properties we define the Hubble parameters
\begin{equation}
 \vec{H}= \frac{d}{dt}\vec{\alpha}
\label{andre12}
\end{equation}
where $t$ is the comoving time. From \eqref{burt15} and \eqref{burt17}
we easily find
\begin{equation}
 \vec{H}=\frac{\vec{p}}{t-t_0}\; ,\qquad 
  \vec{p}= \left(
                \begin{array}{c}
                p \\
                q \\
                \frac{1}{6}q \\
                \end{array}
                \right)\; .
 \label{Hubble}
\end{equation}
Note that the powers $\vec{p}$ satisfy the constraints
\begin{equation}
 \vec{p}^{T}G\vec{p}=0 \qquad  \vec{d}\cdot \vec{p}=1\; .
\label{hello2}
\end{equation}
These relations are characteristic for rolling radii solutions~\cite{mueller}
which are fundamental cosmological solutions of weakly coupled heterotic
string theory. Comparison of the equations of motion~\eqref{burt22}
indeed shows that the scale factors $\vec{\a}$ behave like rolling
radii. The original rolling radii solutions describe freely evolving scale
factors of a product of homogeneous, isotropic spaces.
In our case, the scale factors also evolve freely (since the time-dependent
part of the equations of motion, obtained after separating variables, does
not contain a potential) but they describe the time evolution of the
domain wall. This also proves our earlier claim that the potential terms
in the five-dimensional action~\eqref{S5} do not directly influence the
time-dependence but are canceled by the static domain wall part of the
solution.

Let us now be more specific about the cosmological properties of our
solution. First note from eq.~\eqref{Hubble} that there exist two
different types of time ranges, namely $t<t_0$ and $t>t_0$. In the first
case, which we call the $(-)$ branch, the evolution starts at
$t\rightarrow -\infty$ and runs into a future curvature
singularity~\cite{lowc1,lowc2} at
$t=t_0$. In the second case, called the $(+)$ branch, we start out in
a past curvature singularity at $t=t_0$ and evolve toward
$t\rightarrow\infty$. In summary, we therefore have the branches
\begin{equation}
 t\in \left\{\begin{array}{cc}
             \mbox{[$-\infty,t_{0}$]} & (-)\;\mbox{branch} \\
             \mbox{[$t_{0},+\infty$]} & (+)\;\mbox{branch} \\
             \end{array}
             \right. \; .
\label{andre11}
\end{equation}
For both of these branches we have two options for the powers
$\vec{p}$, defined in eq.~\eqref{Hubble}, corresponding to the two
different signs in eq.~\eqref{burt16}. Numerically, we find
\begin{equation}
 \vec{p}_{\uparrow}\simeq \left(
                \begin{array}{c}
                +.06 \\
                +.81 \\
                +.14 \\
                \end{array}
                \right)\; ,\qquad
 \vec{p}_{\downarrow}\simeq \left(
                \begin{array}{c}
                +.48 \\
                -.45 \\
                -.08 \\
                \end{array}
                \right)
\label{andre16}
\end{equation}
for the upper and lower sign in~\eqref{burt16} respectively. We recall
that the three entries in these vectors specify the evolution powers for
the spatial worldvolume of the three-brane, the domain wall separation and
the Calabi-Yau volume. The expansion of the domain wall worldvolume has
so far been measured in terms of the five-dimensional Einstein frame
metric $g^{(5)}_{\m\n}$. This is also what the above numbers $p_1$ reflect.
Alternatively, one could measure this expansion
with the four-dimensional Einstein frame metric $g^{(4)}_{\m\n}$ so
that the curvature scalar on the worldvolume is canonically normalized.
{}From the relation
\begin{equation}
 g^{(4)}_{\m\n} = \left( g_{11,11}\right)^{1/2}g^{(5)}_{\m\n}
\end{equation}
we find that this modifies $p_1$ to
\begin{equation}
 \tilde{p}_1 = p_1+\frac{p_2}{2}\; .\label{conv}
\end{equation}
In the following, we will discuss both frames. We recall that the separation
condition $\b =\g$ implies that the internal Calabi-Yau space always
tracks the orbifold. In the discussion we can, therefore, concentrate on
the spatial worldvolume and the orbifold, corresponding to the first
and second entries in \eqref{andre16}. Let us first consider the
$(-)$ branch. In this branch $t\in [-\infty ,t_0]$ and, hence, $t-t_0$ is
always negative. It follows from eq.~\eqref{Hubble} that a subspace will
expand if its $\vec{p}$ component is negative and contract if it is
positive. For the first set of powers $\vec{p}_{\uparrow}$ in
eq.~\eqref{andre16} both the worldvolume and the orbifold contract in
the five-dimensional Einstein frame. The same conclusion holds in the
four-dimensional Einstein frame. For the second set, $\vec{p}_{\downarrow}$,
in both frames the worldvolume contracts while the orbifold expands.
Furthermore, since the Hubble parameter of the orbifold increases in time
the orbifold undergoes superinflation.

Now we turn to the $(+)$ branch. In this branch $t\in [t_0,\infty ]$ and,
hence, $t-t_0$ is always positive. Consequently, a subspace expands for
a positive component of $\vec{p}$ and contracts otherwise. In addition,
since the absolute values of all powers $\vec{p}$ are smaller than one,
an expansion is always subluminal. For the vector $\vec{p}_{\uparrow}$
the worldvolume and the orbifold expand in both frames. On the other
hand, the vector $\vec{p}_{\downarrow}$ describes an expanding worldvolume
and a contracting orbifold in both frames. This last solution perhaps
corresponds most closely to our notion of the early universe.

Let us briefly summarize the basic geometry and physics of the solutions
discussed in this section, emphasizing the $(+)$ branch,
$\vec{p}_{\downarrow}$ solution for concreteness. We find that at the GUT
scale of around $10^{16}$ GeV, Ho\v rava-Witten theory compactifies on a
Calabi--Yau three--fold to become an effective five--dimensional theory with
two four--dimensional boundaries. Anomaly cancellation constrains this
effective theory to be a specific gauged form of $N=1$ supergravity coupled to
hyper and vector supermultiplets in the bulk with associated $N=1$ gauge
theories on each boundary. The size of the fifth--dimension can be anywhere
between four or five times the inverse GUT scale to much larger, perhaps 
inverse $10^{14}$
Gev. This theory does not admit flat space as a solution. Rather, the static
vacuum is found to be a pair of BPS three-branes each located on one of the
orbifold fixed planes. There are natural cosmological solutions associated
with this vacuum state. These solutions expand or contract in the
four--dimensional directions of the low energy Universe as well as expanding
or contracting in the fifth, orbifold direction. Among a plethora of
possibilities mentioned above, one solution, the $(+)$ branch with 
$\vec{p}_{\downarrow}$, corresponds to subluminal
expansion of real four--dimensional spacetime while the associated
fifth--dimension contracts. One assumes that other, perhaps non-perturbative,
physics will stabilize the moduli of the fifth--direction. If so, this
solution would be a candidate for a Robertson-Walker expanding phase of
the early Universe.


\section*{Lecture 3:}

In this lecture, we explore cosmological solutions with non-vanishing
Ramond-Ramond form backgrounds. Solution of these cosmological scenarios is
mathematically intricate and requires the use of methods, related to Toda
theory, introduced in our previous work. We present one complete example with
a single Ramond-Ramond scalar turned on and discuss its physical
interpretation in detail.

\section*{Cosmological solutions with Ramond forms}


Thus far, we have looked for both static and cosmological solutions where the
form fields $\x, {\cal{A}}_{\a}$ and $\s$ have been set to zero.
As discussed in previous papers~\cite{lowc1,lowc2}, turning on one or
several such fields can drastically alter the solutions and their cosmological
properties. Hence, we would like to explore cosmological solutions with
such non-trivial fields. For clarity, in this lecture we will restrict the
discussion to turning on the Ramond-Ramond scalar $\x$ only.

The Ansatz we will use is the following. For the metric and dilaton field, we 
choose
\bea 
 ds_5^2 &=& -N(\tau,y)^{2}d\tau^{2}+a(\tau,y)^2dx^{m} dx^{n}\eta_{mn}+
            b(\tau,y)^2dy^2   \label{dan1}\\
 V &=& V(\tau,y)\; .\nn
\eea
For the $\x$ field, we assume that $\x=\x(\tau,y)$ and, hence, the field 
strength $F_{\a}=\partial_{\a}\xi$ is given by
\begin{equation}
 F_{0}=Y(\tau,y)\; ,\qquad  F_{5}=X(\tau,y)\; .
\label{dan2}
\end{equation}
All other components of $F_{\a}$ vanish. Note that since $\x$ is complex, both 
$X$ and $Y$ are complex. Once again, we will solve the equations of motion
by separation of variables. That is, we let 
\bea
 N(\tau,y)=n(\tau)N(y) \nn\\
 a(\tau,y)=\alpha(\tau)a(y) \nn\\
 b(\tau,y)=\beta(\tau)b(y) \label{dan3}\\
 V(\tau,y)=\gamma(\tau)V(y)\nn
\eea
and
\bea
 X(\tau,y)=\chi(\tau)X(y)\\
 Y(\tau,y)=\phi(\tau)Y(y)\; .
\label{dan4}
\eea
Note that, in addition to the $\x$ field, we have also allowed for
the possibility that $N(y)\neq a(y)$. Again, there is no a priori reason to 
believe that a solution can be found by separation of variables. However, as 
above, there is indeed such a solution, although the constraints required to 
separate variables are more subtle. It is instructive to present one of the 
equations of motion. With the above Ansatz, the $g_{00}$ equation of motion 
becomes~\footnote{In the following, $N$, $a$, $b$, $V$ denote the
$y$-dependent part of the Ansatz~\eqref{dan3}.}
\begin{multline}
\label{dan5}
 \frac{N^{2}}{b^{2}}\left(\frac{a''}{a}+\frac{{a'}^{2}}{a^{2}}-\frac{a'}{a}
  \frac{b'}{b}+\frac{1}{12}\frac{{V'}^{2}}{V^{2}}
  +\frac{\alpha_{0}^{2}}{18}\frac{b^{2}}{V^{2}}\frac{\beta^{2}}{\gamma^{2}}
  -\frac{\sqrt{2}\alpha_{0}}{3}
     \frac{b}{V}(\delta(y)-\delta(y-\pi\rho))\frac{\beta}{\gamma}
  \right) = \\
  \frac{\beta^{2}}{n^{2}}\left(\frac{\dot{\alpha}^{2}}{\alpha^{2}}
  +\frac{\dot{\alpha}}{\alpha}\frac{\dot{\beta}}{\beta}-\frac{1}{12}
  \frac{\dot{\gamma}^{2}}{\gamma^{2}}\right)
  -\frac{N^{2}}{3b^{2}}\frac{|X|^{2}}{V}\frac{|\c |^{2}}{\gamma}
  -\frac{\beta^{2}}{3n^{2}}\frac{|Y|^{2}}{V}\frac{|\phi |^{2}}{\gamma}
\end{multline}
Note that if we set $X=Y=0$ and $N=a$ this equation becomes identical to
\eqref{burt7}. We now see that there are two different types of obstructions 
to the separation of variables. The first type, which we encountered in the 
previous section, is in the two terms proportional to $\alpha_{0}$. 
Clearly, we can separate variables only if we demand that
\bea
\beta=\gamma
\label{dan6}
\eea
as we did previously. However, for non-vanishing $X$ and $Y$ this is not 
sufficient. The problem, of course, comes from the last two terms in \eqref
{dan5}. There are a number of options one could try in order to separate 
 variables in these terms. It is important to note that $X$ and $Y$ are not 
completely independent, but are related to each other by the integrability 
condition  $\partial_{\tau}X(\tau,y)=\partial_{y}Y(\tau,y)$. We find that, 
because of this condition, it is impossible to obtain a solution by 
separation of variables that has both $X(\tau,y)$ and $Y(\tau,y)$ 
non-vanishing. Now $X(\tau,y)$, but not $Y(\tau,y)$, can be made to vanish 
by taking $\x=\x(\tau)$; that is, $\x$ is a function of $\tau$ only. However
we can find no solution by separation of variables under this circumstance. 
Thus, we are finally led to the choice $\x=\x(y)$. In this case 
$Y(\tau,y)=0$ and we can, without lose of generality, choose
\bea
 \chi=1\; .
\label{dan7}
\eea
At this point, the only obstruction to separation of variables in equation
\eqref{dan5} is the next to last term, $N^2|X|^2/3b^2V\g$. Setting
$\gamma=\mbox{const}$ is too restrictive, so we must demand that 
\bea
  X = \frac{bV^{1/2}}{N}c_{0}e^{i\theta(y)}
\label{dan8}
\eea
where $c_{0}$ is a non-zero but otherwise arbitrary real constant and
$\theta(y)$ is an, as yet, undetermined phase.
Putting this condition into the $\x$ equation of motion
\bea
 \partial_{y}\left(\frac{a^3N}{bV}X\right)=0
\label{dan9}
\eea
we find that $\theta$ is a constant $\theta_0$ and $a\propto
V^{\frac{1}{6}}$ with arbitrary coefficient. Note that the last condition is 
consistent with the static vacuum solution~\eqref{final3}. Inserting this 
result into the $g_{05}$ equation of motion
\bea
 \frac{\dot{\alpha}}{\alpha}\left(\frac{a'}{a}-\frac{N'}{N}\right)=
 \frac{\dot{\beta}}{\beta}\left(\frac{a'}{a}-\frac{1}{6}\frac{V'}{V}\right)
\label{dan10}
\eea 
we learn that $N\propto a$ with arbitrary coefficient. Henceforth, we choose
$N=a$ which is consistent with the static vacuum solution~\eqref{final3}.
Inserting all of these results, the $g_{00}$ equation of motion now becomes
\begin{multline}
 \label{dan11}
 \frac{a^{2}}{b^{2}}\left(\frac{a''}{a}+\frac{{a'}^{2}}{a^{2}}-\frac{a'}{a}
  \frac{b'}{b}+\frac{1}{12}\frac{{V'}^{2}}{V^{2}}+\frac{\alpha_{0}^{2}}
   {18}\frac{b^{2}}{V^{2}}-\frac{\sqrt{2}}{3}
   \alpha_{0}\frac{b}{V}(\delta(y)-\delta(y-\pi\rho))\right) = \\
\frac{\beta^{2}}{n^{2}}\left(\frac{\dot{\alpha}^{2}}{\alpha^{2}}
 +\frac{\dot{\alpha}}{\alpha}\frac{\dot{\beta}}{\beta}-\frac{1}{12}
  \frac{\dot{\gamma}^{2}}{\gamma^{2}}\right)-\frac{c_{0}^{2}}{3}
  \frac{1}{\gamma}
\end{multline}
Note that the left hand side is of the same form as the static vacuum equation 
\eqref{burt3}. The effect of turning on the $\x$ background is to add a purely
$\tau$ dependent piece to the right hand side.
Putting these results into the remaining four equations of motion, we 
find that they too separate, with the left hand side being purely $y$ 
dependent and the right hand side purely $\tau$ dependent. Again, we 
find that in these equations the left hand sides are identical to those 
in the static vacuum equations and the effect of turning on $\x$ is to add
extra $\tau$ dependent terms to the right hand sides. In each equation,
both sides must now equal the 
same constant which, for simplicity, we take to be zero. The $y$ equations 
for $a$, $b$ and $V$ thus obtained by setting the left hand side to zero
are identical to the static vacuum equations. Hence, we have shown that
\bea
 N=a &=&a_0H^{1/2}\nn \\
 b &=& b_0H^2\qquad\qquad H=\frac{\sqrt{2}}{3}\a_{0}|y|+h_0 \label{dan12}\\
 V &=&b_0H^3 \nn \\
 X &=& x_{0}H^{3} \nn
\eea
where $x_{0}=c_{0}e^{i\theta_0}a_{0}^{-1}b_{0}^{3/2}$ is an arbitrary
constant. 

The $\tau$ equations obtained by setting the right hand side to zero are the 
following.
\bea
 \frac{\dot{\alpha}^{2}}{\alpha^{2}}
 +\frac{\dot{\alpha}}{\alpha}\frac{\dot{\beta}}{\beta}-\frac{1}{12}
  \frac{\dot{\gamma}^{2}}{\gamma^{2}}-\frac{c_{0}^{2}}{3}
  \frac{n^{2}}{\beta^{2}\gamma}=0
\label{dan13}
\eea
\bea
 2\frac{\ddot{\alpha}}{\alpha}-2\frac{\dot{\alpha}}{\alpha}\frac{\dot{n}}{n}
  +\frac{\ddot{\beta}}{\beta}-\frac{\dot{\beta}}{\beta}\frac{\dot{n}}{n}
   +\frac{\dot{\alpha}^{2}}{\alpha^{2}}+2\frac{\dot{\alpha}}{\alpha}
   \frac{\dot{\beta}}{\beta}+\frac{1}{4}\frac{\dot{\gamma}^{2}}
  {\gamma^{2}}-c_{0}^{2}\frac{n^{2}}{\beta^{2}\gamma} =0
\label{dan14}
\eea
\bea
 \frac{\ddot{\alpha}}{\alpha}-\frac{\dot{\alpha}}{\alpha}\frac{\dot{n}}{n}
  +\frac{\dot{\alpha}^{2}}{\alpha^{2}}+\frac{1}{12}\frac{\dot{\gamma}^{2}}
  {\gamma^{2}}+\frac{c_{0}^{2}}{3}\frac{n^{2}}{\beta^{2}\gamma} =0
\label{dan15}
\eea
\bea
 \frac{\ddot{\gamma}}{\gamma}+3\frac{\dot{\a}\dot{\g}}{\a\g}+
 \frac{\dot{\b}\dot{\g}}{\b\g}-\frac{\dot{\g}^2}{\g^2}
  -\frac{\dot{n}\dot{\g}}{n\g}
  -2c_{0}^{2}\frac{n^{2}}{\beta^{2}\dot{\gamma}} =0
\label{dan16}
\eea
In these equations we have, once again, displayed $\beta$ and $\gamma$ 
independently, although they should be solved subject to the 
condition $\beta=\gamma$. Note that the above 
equations are similar to the $\tau$ equations in the previous section, 
but each now has an additional term proportional to $c_{0}^{2}$. These 
extra terms considerably complicate finding a solution of the $\tau$ 
equations. Here, however, is where the formalism introduced in the 
previous section becomes important. Defining $\hat{\alpha},\hat{\beta}$ 
and $\hat{\g}$ as in \eqref{burt17}, and $\vec{\alpha}, E$ and $G$
as in \eqref{burt18}, \eqref{burt19} and \eqref{burt20} respectively,
the equations \eqref{dan13}-\eqref{dan16} can be written in the form
\begin{equation}
 \frac{1}{2}E\dot{\vec{\alpha}}^{T}G\dot{\vec{\alpha}}+E^{-1}U=0\; ,\qquad
 \frac{d}{d\t}\left(EG\dot{\vec{\alpha}}\right)+E^{-1}\frac{\partial U}
 {\partial\vec{\alpha}}=0
\label{dan18}
\end{equation}
where the potential $U$ is defined as
\begin{equation}
 U=2c_{0}^{2}e^{\vec{q}\cdot\vec{\alpha}}
\label{dan19}
\end{equation}
with
\begin{equation}
 \vec{q}= \left(
                \begin{array}{c}
                6 \\
                0 \\
                -6 \\
                \end{array}
                \right)\; .
\label{dan20}
\end{equation}
We can now exploit the gauge freedom of $n$ to simplify these equations.
Choose the gauge
\begin{equation}
 n=e^{(\vec{d}-\vec{q})\cdot\vec{\alpha}}
\label{dan21}
\end{equation}
where $\vec{d}$ is defined in \eqref{burt18}. Then
$E$ becomes proportional to the potential $U$ so that the potential
terms in \eqref{dan18} turn into constants. Thanks to this simplification,
the equations of motion can be integrated which leads to the
general solution~\cite{lowc1,lowc2}
\bea
 \vec{\alpha}=\vec{c}\,\ln|\tau_{1}-\tau|+\vec{w}\,\ln\left(\frac{s\tau}
 {\tau_{1}-\tau}\right) +\vec{k}
\label{dan24}
\eea
where $\tau_{1}$ is an arbitrary parameter which we take, without loss of 
generality, to be positive and
\begin{equation}
 \vec{c}=2\frac{G^{-1}\vec{q}}{<\vec{q},\vec{q}>}\; ,\qquad
 s=\mbox{sign}(<\vec{q},\vec{q}>)\; .
\label{dan25}
\end{equation}
The scalar product is defined as
$<\vec{q},\vec{q}>=\vec{q}^{T}G^{-1}\vec{q}$. The vectors $\vec{w}$ and 
$\vec{k}$ are integration constants subject to the constraints
\bea
 \vec{q}\cdot \vec{w}&=&1 \nn \\
 \vec{w}^{T}G\vec{w}&=&0 \label{dan25p}\\
 \vec{q}\cdot \vec{k}&=& \ln\left(c_{0}^{2}|<\vec{q},\vec{q}>|\right) \nn
\eea
This solution is quite general in that it describes an arbitrary number
of scale factors with equations of motion given by~\eqref{dan18}. Let us now
specify to our example. For $G$ and $\vec{q}$ as given in eq.~\eqref{burt20}
and \eqref{dan20} we find that
\begin{equation}
 <\vec{q},\vec{q}>=1
\label{dan26}
\end{equation}
hence $s=1$, and
\begin{equation}
 \vec{c}= \left(
                \begin{array}{c}
                0 \\
                -2 \\
                -\frac{1}{3} \\
                \end{array}
                \right)\; .
\label{dan27}
\end{equation}
Recall that we must, in addition, demand that $\beta=\gamma$. Note that the 
last two components of $\vec{c}$ are consistent with this equality.
We can also solve the constraints~\eqref{dan25p} subject to the
condition $\beta=\gamma$. The result is
\begin{equation}
 \vec{w}=     \left(
                \begin{array}{c}
                w_{3}+\frac{1}{6} \\
                6w_{3} \\
                w_{3} \\
                \end{array}
                \right)\; ,\qquad
 \vec{k}=   \left(
                \begin{array}{c}
                k_{3}+\frac{1}{6}\ln c_{0}^{2} \\
                6k_{3} \\
                k_{3} \\
                \end{array}
                \right)  
\label{dan29}
\end{equation}
where
\begin{equation}
 w_{3}=-\frac{1}{6}\pm\frac{\sqrt{3}}{12}
\end{equation}
and $k_{3}$ is arbitrary. We conclude that in the gauge specified by 
\eqref{dan21}, the solution is given by
\bea
 \hat{\alpha}=(w_{3}+\frac{1}{6})\ln\left(\frac{\tau}{\tau_{1}-\tau}\right)
 +k_{3}+\frac{1}{6}\ln c_{0}^{2} \nn\\
 \hat{\beta}= -2\ln |\tau_{1}-\tau| +6w_{3}\ln\left(\frac{\tau}
 {\tau_{1}-\tau}\right)+6k_{3} \label{dan30}\\
 \hat{\gamma}=-\frac{1}{3}\ln |\tau_{1}-\tau|+w_{3}\ln\left(\frac{\tau}
 {\tau_{1}-\tau}\right)+k_{3}\nn
\eea
with $w_{3}$ as above. As a consequence of $s=1$, the range for $\t$
is restricted to
\begin{equation}
 0<\tau<\tau_{1}
\label{dan34}
\end{equation}
in this solution. Let us now summarize our result. We have found a
cosmological solution with a nontrivial Ramond-Ramond scalar $\x$
starting with the separation Ansatz~\eqref{dan3}. To achieve separation
of variables we had to demand that $\b =\g$, as previously, and
that the Ramond-Ramond scalar depends on the orbifold coordinate but not on
time. Then the orbifold dependent part of the solution is given by
eq.~\eqref{dan12} and is identical to the static domain wall solution
with the addition of the Ramond-Ramond scalar. The time dependent
part, in the gauge~\eqref{dan21}, is specified by eq.~\eqref{dan30}.
Furthermore, we have found that the time-dependent part of the equations
of motion can be cast in a form familiar from cosmological solutions
studied previously~\cite{lowc1,lowc2}.
Those solutions describe the evolution for
scale factors of homogeneous, isotropic subspaces in the presence of
antisymmetric tensor fields and are, therefore, natural generalizations of
the rolling radii solutions. Each antisymmetric tensor field introduces
an exponential type potential similar to the one in eq.~\eqref{dan19}. For
the case with only one nontrivial form field, the general solution could
be found and is given by eq.~\eqref{dan24}. We have, therefore, constructed
a strong coupling version of these generalized rolling radii solutions
with a one-form field strength, where the radii now specify the domain wall 
geometry rather that the size of maximally symmetric subspaces. We stress
that the potential $U$ in the time-dependent equations of motion does
not originate from the potentials in the action~\eqref{S5} but from
the nontrivial Ramond-Ramond scalar. The potentials in the action are
canceled by the static domain wall part of the solution, as in the
previous example. 

{}From the similarity to the known generalized rolling
radii solutions, we can also directly infer some of the basic
cosmological properties of our solution, using the results of
ref.~\cite{lowc1,lowc2}. We expect the integration constants to split
into two disjunct sets which lead to solutions in the $(-)$ branch,
comoving time range $t\in [-\infty ,t_0]$, and the $(+)$ branch,
comoving time range $t\in [t_0,\infty ]$, respectively. The $(-)$ branch
ends in a future curvature singularity and the $(+)$ branch starts
in a past curvature singularity. In both branches the solutions behave
like rolling radii solutions asymptotically; that is, at
$t\rightarrow -\infty ,t_0$ in the $(-)$ branch and at 
$t\rightarrow t_0,\infty$ in the $(+)$ branch. The two asymptotic
regions in both branches have different expansion properties in general
and the transition between them can be attributed to the nontrivial
form field.

Let us now analyze this in more detail for our solution, following the
method presented in ref.~\cite{lowc1,lowc2}. First we should
express our solution in terms of the comoving time $t$ by integrating
$dt=n(\t )d\t$. The gauge parameter $n(\t )$ is explicitly given by
\begin{equation}
 n=e^{(\vec{d}-\vec{q})\cdot\vec{k}}|\tau_{1}-\tau|^{-x+\Delta-1}|\tau|^{x-1}
\label{dan33}
\end{equation}
where
\begin{equation}
 x=\vec{d}\cdot\vec{w}\; , \qquad\Delta=\vec{d}\cdot\vec{c}\; .
\label{dan31}
\end{equation}
Given this expression, the integration cannot easily be performed in general
except in the asymptotic regions $\t\rightarrow 0,\t_1$. These regions
will turn out to be precisely the asymptotic rolling-radii limits. Therefore,
for our purpose, it suffices to concentrate on those regions.
Eq.~\eqref{dan33} shows that the resulting range for the comoving time
depends on the magnitude of $\Delta$ and $x$ (note that $\Delta$ is a fixed
number, for a given model, whereas $x$ depends on the integration constants).
It turns out that for all values of the integration constants we have either
$x<\Delta$ or $x>0>\Delta$. This splits the space of integration
constant into two disjunct sets corresponding to the $(-)$ and the
$(+)$ branch as explained before. More precisely, we have the mapping
\begin{equation}
 \t\rightarrow t\in\left\{\begin{array}{clll}
       \left[ -\infty ,t_0\right]&{\rm for}\; x<\D<0\; ,&(-)\;{\rm branch} \\
       \left[ t_0,+\infty\right]&{\rm for}\; x>0>\D\; ,&(+)\;{\rm branch}
       \end{array}\right.
\label{mapping}
\end{equation}
where $t_0$ is a finite arbitrary time (which can be different for the
two branches). We recall that the range of $\t$ is $0<\t <\t_1$. The
above result can be easily read off from the expression~\eqref{dan33}
for the gauge parameter. Performing the integration in the asymptotic
region we can express $\t$ in terms of the comoving time and find
the Hubble parameters, defined by eq.~\eqref{andre12}, and the
powers $\vec{p}$. Generally, we have
\begin{equation}
 \vec{p} = \left\{\begin{array}{cll} \frac{\vec{w}}{x}&{\rm at}&\t\simeq 0 \\
                       \frac{\vec{w} -\vec{c}}{x-\D}&{\rm at}&\t\simeq\t_1
       \end{array}\right. \; . \label{p_expr}
\end{equation}
Note that, from the mapping~\eqref{mapping}, the expression at $\t\simeq 0$
describes the evolution powers at $t\rightarrow -\infty$ in the $(-)$
branch and at $t\simeq t_0$ in the $(+)$ branch; that is, the evolution
powers in the early asymptotic region. Correspondingly, the expression
for $\t\simeq\t_1$ applies to the late asymptotic regions; that is,
to $t\simeq t_0$ in the $(-)$ branch and to $t\rightarrow\infty$ in the
$(+)$ branch. As before, these powers satisfy the rolling
radii constraints~\eqref{hello2}.

Let us now insert the explicit expression for $\vec{d}$, $\vec{w}$ and
$\vec{c}$, eqs.~\eqref{burt18},\eqref{dan29} and \eqref{dan27}, that specify
our example into those formulae. First, from eq.~\eqref{dan31} we find that
\begin{equation}
 x=-1\pm3\frac{\sqrt{3}}{4}\; ,\qquad
 \Delta=-2\; .
\label{dan32}
\end{equation}
Note that the space of integration constants just consists of two points
in our case, represented by the two signs in the expression for $x$ above.
Clearly, from the criterion~\eqref{mapping} the upper sign leads to
a solution in the $(+)$ branch and the lower sign to a solution in the
$(-)$ branch. In each branch we therefore have a uniquely determined
solution. Using eq.~\eqref{p_expr} we can calculate the asymptotic
evolution powers in the $(-)$ branch
\begin{equation}
 \vec{p}_{-,t\rightarrow -\infty}=\left(
                \begin{array}{c}
                +.06 \\
                +.81 \\
                +.13 \\
                \end{array}
                \right)\; ,\qquad
 \vec{p}_{-,t\rightarrow t_0}=\left(
                \begin{array}{c}
                +.48 \\
                -.45 \\
                -.08 \\
                \end{array}
                \right)\; .
\label{kelly14}
\end{equation}
Correspondingly, for the $(+)$ branch we have
\begin{equation}
 \vec{p}_{+,t\rightarrow t_0}=\left(
                \begin{array}{c}
                +.48 \\
                -.45 \\
                -.08 \\
                \end{array}
                \right)\; ,\qquad
 \vec{p}_{+,t\rightarrow\infty}=\left(
                \begin{array}{c}
                +.06 \\
                +.81 \\
                +.13 \\
                \end{array}
                \right)\; .
\end{equation}
Note that these vectors are in fact the same as in the $(-)$
branch, with the time order being reversed. This happens because they
are three conditions on the powers $\vec{p}$ that hold in both branches,
namely the two rolling radii constraints~\eqref{hello2} and the separation
constraint $\b =\g$, eq.~\eqref{dan6}, which implies that $p_3=6p_2$.
Since two of these conditions are linear and one is quadratic, we expect
at most two different solutions for $\vec{p}$.
As in the previous solution, the time variation of the Calabi-Yau volume
(third entry) is tracking the orbifold variation (second entry) as a
consequence of the separation condition and, hence, needs not to be
discussed separately. The first entry gives the evolution power for the
spatial worldvolume in the five-dimensional Einstein frame. For a conversion
to the four-dimensional Einstein frame one should again apply
eq.~\eqref{conv}. It is clear from the above numbers, however, that this
conversion does not change the qualitative behaviour of the worldvolume
evolution in any of the cases. Having said this, let us first discuss
the $(-)$ branch. At $t\rightarrow -\infty$ the powers are
positive and, hence, the worldvolume and the orbifold are contracting.
The solution then undergoes the transition induced by the Ramond-Ramond
scalar. Then at $t\simeq t_0$ the worldvolume is still contracting while
the orbifold has turned into superinflating expansion.
In the $(+)$ branch we start out with a subluminally expanding worldvolume
and a contracting orbifold at $t\simeq t_0$. After the transition both
subspaces have turned into subluminal expansion.

To conclude, Ramond forms have a major effect on the behaviour of cosmological
solutions in five--dimensional Ho\v rava-Witten theory. Generically, they
change the coefficients governing the asymptotic behaviour of these solutions
in a given branch at very early and very late times. This allows the orbifold
to, say, start out contracting and then, at some intermediate time to halt,
turn around and superluminally expand. For the moment, it is not clear how
such solutions fit into a theory of the real Universe. However, they open the
door to new and novel mechanisms in early Universe cosmology which may someday
be necessary to explain observable phenomenon.

\section*{Conclusion}

In these lectures we have presented examples of cosmological
solutions in five-dimensional Ho\v rava-Witten theory. They are
physically relevant in that they are related to the exact BPS three--brane
pair in five dimensions, whose $d=4$ worldvolume theory exhibits $N=1$
supersymmetry. A wider class of such cosmological solutions can be
obtained and will be presented elsewhere. We expect solutions
of this type to provide the fundamental scaffolding for theories of the
early universe derived from Ho\v rava-Witten theory, but they are clearly
not sufficient as they stand. The most notable deficiency is the fact
that they are vacuum solutions, devoid of any matter, radiation or
potential stress-energy. Inclusion of such stress-energy is essential
to understand the behaviour of early universe cosmology. A study of its
effect on the cosmology of Ho\v rava-Witten theory is presently 
underway~\cite{lownext}.

\vspace{0.4cm}

{\bf Acknowledgments} 
A.~L.~is supported in part by the European Community under contract No. 
FMRXCT 960090. B.~A.~O.~is supported in part by 
DOE under contract No. DE-AC02-76-ER-03071. D.~W.~is supported in part by
DOE under contract No. DE-FG02-91ER40671.



\end{document}